\documentclass[%
 aip,
 amsmath,amssymb,
 reprint,%
]{revtex4-1}

\usepackage{graphicx}
\usepackage{dcolumn}
\usepackage{bm}

\usepackage[utf8]{inputenc}
\usepackage[T1]{fontenc}
\usepackage{mathptmx}
\usepackage{etoolbox}

\usepackage{siunitx}
\usepackage{multirow}

\makeatletter
\def\@email#1#2{%
 \endgroup
 \patchcmd{\titleblock@produce}
  {\frontmatter@RRAPformat}
  {\frontmatter@RRAPformat{\produce@RRAP{*#1\href{mailto:#2}{#2}}}\frontmatter@RRAPformat}
  {}{}
}%
\makeatother
\begin{document}

\preprint{AIP/123-QED}

\title[ML potentials with explicit dispersion interaction]{Moment Tensor Potential and Equivariant Tensor Network Potential with explicit dispersion interactions}
\author{Olga Chalykh}
\email{olga.chalykh@skoltech.ru}
\affiliation{Skolkovo Institute of
Science and Technology, Skolkovo Innovation Center, Bolshoy boulevard 30, Moscow, 143026, Russian Federation}

\author{Dmitry Korogod}
\affiliation{Skolkovo Institute of
Science and Technology, Skolkovo Innovation Center, Bolshoy boulevard 30, Moscow, 143026, Russian Federation}
\affiliation{Moscow Institute of Physics and Technology, Institutsky lane 9, Dolgoprudny, Moscow region, 141700, Russian Federation}

\author{Ivan S. Novikov}
\affiliation{Skolkovo Institute of
Science and Technology, Skolkovo Innovation Center, Bolshoy boulevard 30, Moscow, 143026, Russian Federation}
\affiliation{HSE University,         Faculty of Computer Science, Pokrovsky boulevard 11, Moscow, 109028, Russian Federation}
\affiliation{Moscow Institute of Physics and Technology, Institutsky lane 9, Dolgoprudny, Moscow region, 141700, Russian Federation}
\affiliation{Emanuel Institute of Biochemical Physics of the Russian Academy of Sciences, 4 Kosygin Street, Moscow, 119334, Russian Federation}

\author{Max Hodapp}
\affiliation{Materials Center Leoben Forschung GmbH (MCL), Austria}

\author{Nikita Rybin}
\affiliation{Skolkovo Institute of
Science and Technology, Skolkovo Innovation Center, Bolshoy boulevard 30, Moscow, 143026, Russian Federation}
\affiliation{Digital Materials LLC, Odintsovo, Kutuzovskaya str. 4A Moscow region, 143001, Russian Federation}

\author{Alexander V. Shapeev}
\affiliation{Skolkovo Institute of
Science and Technology, Skolkovo Innovation Center, Bolshoy boulevard 30, Moscow, 143026, Russian Federation}
\affiliation{Digital Materials LLC, Odintsovo, Kutuzovskaya str. 4A Moscow region, 143001, Russian Federation}

\date{\today}

\newcommand{\rem}[1]{\textsc{\color{red}#1}}

\begin{abstract}

In this study, we investigate the effect of incorporating explicit dispersion interactions in the functional form of machine learning interatomic potentials (MLIPs), particularly in the Moment Tensor Potential and Equivariant Tensor Network potential for an accurate modeling of liquid carbon tetrachloride, methane, and toluene. We show that explicit incorporation of dispersion interactions via D2 and D3 corrections significantly improves the accuracy of MLIPs when the cutoff radius is set to the commonly used value of 5--6~\AA. We also show that for carbon tetrachloride and methane, a substantial improvement in accuracy can be achieved by extending the cutoff radius to 7.5~\AA. However, for accurate modeling of toluene, explicit incorporation of dispersion remains important. Furthermore, we find that MLIPs incorporating dispersion interactions via D2 reach a close level of accuracy to those incorporating D3, implying that D2 is suitable for an accurate modeling of the systems in the study, while being less computationally expensive. We benchmarked the accuracy of the MLIPs on dimer binding curves compared to \textit{ab initio} data and on predicting density and radial distribution functions compared to experiments.

\end{abstract}

\maketitle

\section{\label{sec:Introduction}Introduction}

Machine learning interatomic potentials (MLIPs) have emerged as widely recognized tools for modeling large systems over extended length- and timescales with the accuracy of \textit{ab initio} methods~\cite{Zuo2020,Kozinsky2023,zhang2025roadmap}. This makes MLIPs an attractive choice for simulating systems that are, on the one hand, computationally prohibitive for conventional \textit{ab initio} methods and, on the other hand, beyond the accuracy of (semi-)empirical force fields.

Despite significant advances in the development of MLIPs, several key challenges remain. The primary challenge for MLIPs arises from their locality, which is a consequence of the introduction of a cutoff radius that restricts the number of interacting atoms~\cite{zhang2025roadmap}. Although this approximation works well for metals and alloys, it requires greater caution for low-dimensional materials, materials with defects, or molecules~\cite{babaei2024,behler2021machine,anstine2023}. Moreover, the locality of MLIPs restricts them from accurately capturing long-range interactions, such as electrostatic, induction, and dispersion~\cite{anstine2023}, interactions.

MLIPs that explicitly incorporate long-range interactions have already been presented in the literature. Such MLIPs include PhysNet~\cite{unke2019physnet}, AIMNet2~\cite{isayev2025aimnet2}, SpookyNet~\cite{unke2021spookynet,unke2024biomolecular}, TensorMol~\cite{yao2018tensormol}, and ee4G-HDNNP model~\cite{ko2023accurate}, which incorporate both, electrostatic and dispersion interactions. Other MLIPs such as HDNNP potentials~\cite{behler2011third, behler2021fourth}, and the DPLR model~\cite{zhang2022deep} incorporate electrostatic interactions. At the same time, GAP~\cite{bartok2010gaussian} was applied to study systems with strong dispersion interactions~\cite{veit2019, rowe2020carbon, rowe2018graphene, ibragimova2025}. However, it is also possible to model intramolecular and intermolecular interactions using separate potentials~\cite{Wengert2021,gao2022self}.  Additionally, an alternative approach is to account for long-range interactions via global descriptors~\cite{kabylda2023efficient}.

The most common approach to address long-range interactions is to explicitly incorporate physically motivated terms in the total energy expression~\cite{anstine2023}. Behler proposed a classification of MLIPs into four generations based on how long-range interactions are incorporated~\cite{behler2021,behler2021machine}. The first and second generations correspond to purely local potentials, while the third generation MLIPs explicitly include long-range interactions, with their strength defined by local features such as atomic charges, dependent on the local environment. The fourth generation accounts for the nonlocality of parameters defining long-range interactions, particularly by making charges dependent on the nonlocal environment. This classification is widely used in the context of MLIPs that incorporate electrostatic interactions, and can also be applied to classify MLIPs that account for dispersion interactions~\cite{behler2021machine}. 

To explicitly capture dispersion interactions, the most common approach is to utilize one of the dispersion corrections developed for density functional theory (DFT) calculations~\cite{grimme2016, hermann2017first}, such as Tkatchenko-Scheffler (TS)~\cite{tkatchenko2009}, exchange-dipole moment (XDM)~\cite{becke2005}, D3~\cite{grimme2010}, or D2~\cite{grimme2006} models. For electronic density-dependent dispersion correction schemes, such as TS or XDM, machine learning can be used to predict the most computationally demanding parameters, such as atomic moments and volumes in XDM~\cite{tu2023,tu2024}. Further optimization of the computational cost can be achieved by using averaged parameters instead of machine learning: atomic volumes for TS~\cite{veit2019,muhli2021,ibragimova2025} or moments and atomic volumes for XDM~\cite{tu2023}. The latter approach has been shown to have an accuracy close to that of the scheme using machine-learned coefficients~\cite{tu2023}. The D2 or D3 dispersion corrections, which do not require electronic density for evaluation, can be applied on top of a local potential~\cite{wang2020,ying2023,unke2019physnet,yao2018tensormol}, without machine learning of the dispersion correction parameters. An alternative involves parameterizing the Lennard-Jones potential based on relevant configurations, such as the exfoliation curve for black phosphorus or graphene~\cite{rowe2018graphene,deringer2020}.

Another challenge is the efficient generation of high-quality \textit{ab initio} training sets, which is crucial to ensure accuracy within the training domain, for efficient construction of training set, and to improve the transferability of the resulting potential. To address this challenge, various data-driven approaches have been developed, focusing on optimization of training sets rather than increasing model complexity. These approaches include different variations of active learning (AL), such as methods based on the Query by Committee strategy~\cite{smith2018}, Bayesian inference~\cite{vandermause2020fly}, optimization of entropy of atomic descriptors~\cite{Zapiain2022}, and the D-optimality criterion employed together with the MaxVol algorithm~\cite{Podryabinkin2016}.

In the present work, we focus on the development of MLIPs with explicit treatment of the dispersion interaction, particularly Moment Tensor Potentials (MTPs)~\cite{Shapeev2016} and Equivariant Tensor Network (ETN) potentials~\cite{hodapp2024equivariant}. MTP is implemented in the MLIP-2 package, together with the AL approach based on the D-optimality criterion~\cite{Novikov2021}, which enables efficient training set generation for MTP. ETN is implemented in the MLIP-4 package~\cite{mlip4}. MTP has been widely applied to the modeling of inorganic materials~\cite{korotaev2019, podryabinkin2019accelerating,wang2020ionic}, metals and alloys~\cite{marmolejo2022,tasnadi2021,gil2022,zheng2023multi}, and molten salts~\cite{rybin2024,sun2024interatomic,attarian2022thermophysical}. However, it was not extensively used to study molecular compounds~\cite{rybin2025accelerating,klimanova2024,reicht2024designing,novikov2024oh_hbr}. At the same time, ETN has recently been developed~\cite{hodapp2024equivariant} and has been successfully applied to defects in metallic alloys~\cite{hodapp2025exact}, but has not yet been widely benchmarked in various applications.

Here, our objective is to improve the accuracy and applicability of the MTP and ETN models by explicitly incorporating dispersion interactions via D2 and D3 corrections. In our study, we compare the precision of local MLIPS (MLIPS(l)) with MLIPS with explicit dispersion correction (MLIP+D). We study the accuracy differences between MLIP(l)'s and MLIP+D's when the cutoff radius is set to a value within a commonly used range of 5--6~\AA, thereby constraining the MLIP(l)'s ability to capture the full range of dispersion interactions. We additionally test an alternative approach in which we simply extend the cutoff radius for MLIP(l)'s to 7.5~\AA, allowing them to describe the major part of dispersion interactions, and compare the accuracy of the resulting MLIP(l)'s with MLIP+D's.

This comparative analysis was performed on liquid carbon tetrachloride, methane, and toluene simulations, since these systems exhibit predominant dispersion interactions and are widely represented in the literature. Moreover, carbon tetrachloride and toluene are common nonpolar solvents.  To examine the predictive power of potentials, we calculate the binding curves and compare them with the ones obtained with \textit{ab initio} calculations. We also calculate the density and radial distribution function (RDF) with the fitted MLIPs and compare them to the experimental ones.

This study is devoted to benchmarking the above MLIPs fitted to PBE+D2/D3 on the density, binding curves, and RDF, as the aim of this study is to investigate the change in accuracy of MLIPs after incorporating explicit dispersion interactions in their functional form. We do not consider more complex properties, such as diffusion coefficients, whose accuracy may be affected by the level of theory or quantum nuclear effects, as this investigation is beyond the scope of this work.

\section{\label{sec:Methodology} Methodology}

\subsection{\label{sec:MTPs} Moment Tensor Potential}

Moment Tensor Potential (MTP) is a machine learning interatomic potential which was first proposed for single-component materials~\cite{Shapeev2016} and generalized to multi-component materials~\cite{gubaev2018}. MTP energy $E^{\rm MTP}$ is the sum of contributions $V^{\rm MTP}(\mathfrak{\bm n}_i)$ of atomic neighborhoods ${\bf \mathfrak{n}}_i$ for $N$ atoms
\begin{align} \label{EnergyMTP}
E^{\rm MTP} = \sum \limits_{i=1}^{N} V^{\rm MTP}(\mathfrak{\bm n}_i). 
\end{align}
Each neighborhood is a tuple 
$$\mathfrak{ n}_i = ( \{r_{i1},z_i,z_1 \}, \ldots, \{r_{ij},z_i,z_j \}, \ldots, \{r_{iN_ {\rm nbh}},z_i,z_{N_ {\rm nbh}} \} ),$$ 
where $r_{ij}$ are relative atomic positions, $z_i$, $z_j$ are the types of central and neighboring atoms, and $N_ {\rm nbh}$ is the number of atoms in neighborhood. Each contribution $V^{\rm MTP}(\mathfrak{\bm n}_i)$ in the potential energy $E^{\rm MTP}$ expands through a set of basis functions
\begin{align} \label{SiteEnergyMTP}
V^{\rm MTP}({\bf \mathfrak{n}}_i) = \sum \limits_{\alpha} \xi_{\alpha} B_{\alpha}({\mathfrak{\bm n}}_i),
\end{align} 
where $B_{\alpha}$ are the MTP basis functions, $\xi_{\alpha}$ are the linear parameters to be found. To define the functional form of the MTP basis functions we introduce the so-called moment tensor descriptors:
\begin{equation}\label{MomentTesnsorDescriptors}
M_{\mu,\nu}({\mathfrak{\bm n}}_i)=\sum_{j=1}^{N_{\rm nbh}} f_{\mu}(|r_{ij}|,z_i,z_j) r_{ij}^{\otimes \nu}.
\end{equation}
The descriptor consists of the angular part $r_{ij}^{\otimes \nu}$ (the symbol ``$\otimes$'' denotes the outer product of vectors and, thus, the angular part is the tensor of $\nu$-th order) and the radial part $f_{\mu}(|r_{ij}|,z_i,z_j)$ of the following form:
\begin{align} \label{RadialFunction}
\displaystyle
f_{\mu}(|r_{ij}|,z_i,z_j) = \sum_{\beta} c^{(\beta)}_{\mu, z_i, z_j} T^{(\beta)} (|r_{ij}|) (R_{\rm cut} - |r_{ij}|)^2.
\end{align}
Here $\mu$ is the number of the radial function $f_{\mu}$, $c=\{c^{(\beta)}_{\mu, z_i, z_j}\}$ are the radial parameters to be found, $T^{(\beta)} (|r_{ij}|)$ are polynomial functions, and the term $(R_{\rm cut} - |r_{ij}|)^2$ is introduced to ensure smoothness with respect to the atoms leaving and entering the sphere with the cutoff radius $R_{\rm cut}$.
 
By definition, the MTP basis function $B_{\alpha}$ is a contraction of one or more moment tensor descriptors, yielding a scalar. In order to construct the basis functions $B_{\alpha}$, and, thus, determine a particular functional form of MTP, we define the so-called \emph{level} of moment tensor descriptor:
\begin{equation} \label{eq:LevelMTD}
\displaystyle
{\rm lev} M_{\mu,\nu} = 2 + 4 \mu + \nu.
\end{equation}
We also define the level of MTP basis function:
\begin{equation} \label{LevelMultMTD}
\displaystyle
{\rm lev} B_{\alpha} = \rm {lev} \underbrace {\prod_{p=1}^{P} M_{\mu_p,\nu_p}}_{scalar} = \sum \limits_{p=1}^P (2 + 4 \mu_p + \nu_p).
\end{equation}
A set of MTP basis functions and, thus, a particular functional form of MTP depends on the maximum level, ${\rm lev_{\rm max}}$, which we also call the level of MTP. In the set of MTP basis functions, we include only those with ${\rm lev} B_{\alpha} \leq {\rm lev_{\rm max}}$. 

\subsection{\label{sec:ETNs} Equivariant Tensor Network Potentials}

Polynomial-based potentials, like MTPs, can be generally written as a contraction of a tensor $T$, containing the parameters of the potential, with feature vectors $v$ that contain the same radial and angular information as the moment tensor descriptors \eqref{MomentTesnsorDescriptors} (we will make this precise further below)
\begin{equation}\label{eq:multilinear_potential}
    V({\bf \mathfrak{n}}_i)
    =
    T_{k_1 \ldots k_d}
    v^1_{k_1} \ldots v^d_{k_d}
    ,
\end{equation}
where the dimension $d$ defines the body-order of the potential.
However, Equation \eqref{eq:multilinear_potential} is not convenient because the evaluation scales exponentially with $d\cdot n$, where $n$ is the number of features contained in each of the $v$'s.
For MTPs, this problem is to some extent solved semi-empirically, e.g., by contracting the radial features before entering the moment tensor descriptors.

A way to explicitly control the growth of the potential's complexity is to represent $T$ with a tensor network.
Tensor networks are a way of representing high-dimensional tensors in a low-rank format by factorizing the full tensor into smaller tensors, up to the order of three.
There are different formats, with tensor trains, hierarchical Tucker, or PEPS, arguably being among the most popular ones~\cite{cichocki_tensor_2017,orus_tensor_2019}, but their contractions with vectors $v^1_{k_1}$, $v^2_{k_2}$, etc., can all be realized by sequences of contractions of up-to-order-three tensors
\begin{align*}
    u^1_{k_1'} &= T^1_{k_1' k_1} v^1_{k_1}, \\
    u^2_{k_2'} &= T^2_{k_2' k_2 k_1'} v^2_{k_2} u^1_{k_1'}, \\
    u^3_{k_3'} &= T^3_{k_3' k_3 k_2'} v^3_{k_3} u^2_{k_2'}, \\
          &\ldots
          .
\end{align*}

The main difference between ETNs and conventional tensor networks is the implementation of symmetry constraints that render the ETN invariant under the actions of the corresponding symmetry group.
In our case of interatomic potentials, we require that the energy per atom $V({\bf \mathfrak{n}}_i)$ remains invariant under the actions of the group of rotations SO(3).%
\footnote{To encode the full O(3) invariance (rotations and reflections) into ETNs, it suffices that $V({\bf \mathfrak{n}}_i)$ is a real quantity (more details are provided in the original paper~\cite{hodapp2024equivariant}, section 3.4.2)}
To that end, we require the feature vectors to be SO(3)-equivariant vectors, that rotate correspondingly with a basis change under SO(3).
In the following, we consider a decomposition of each $v$ into an irreducible covariant representation of SO(3) using spherical harmonics.
Thus, each index $k$ of a feature vector $v_k$ is a multi-index $k = (\ell mn)$, with $\ell = 0, \ldots, L$ being the index of the subspace of the irreducible representation, $m \in \{ -\ell, -\ell + 1, \ldots, \ell \}$ being the dimension of the subspace, and $n = 1, \ldots, N(\ell)$ is the number of radial channels corresponding to each $\ell$.
We point out that we intentionally deviate from the ordering $n\ell m$, commonly used in quantum physics, that puts the index $n$ in front of $\ell$ and $m$.
We chose this notation because each multi-index in our tensor network may depend on a different $\ell$. Therefore, since $m$ and $n$ are always assumed to depend on $\ell$, the ordering $\ell m n$ appears to be more comprehensible in our context.

With this definition of the feature vectors, a sufficient condition for $V({\bf \mathfrak{n}}_i)$ being invariant under actions of SO(3) is that the tensors $T$ are equivariant maps of these covariant vectors.
According to the Wigner-Eckhart Theorem, any $T$ with three multi-indices $\{ (\ell_i, m_i, n_i) \}_{i=1,\ldots,3}$, can be factorized to
\[
    T_{(\ell_1 m_1 n_1)(\ell_2 m_2 n_2)(\ell_3 m_2 n_3)}
    =
    \theta_{(\ell_1 n_1)(\ell_2 n_2)(\ell_3 n_3)}
    C_{(\ell_1 m_1)(\ell_2 m_2)(\ell_3 m_3)}
    ,
\]
where $\theta_{(\ell_1 n_1)(\ell_2 n_2)(\ell_3 n_3)}$ is the tensor of model coefficients, and $C_{(\ell_1 m_1)(\ell_2 m_2)(\ell_3 m_3)}$ is the Clebsch-Gordan coefficient that defines the symmetry group.
As a tensor network, we use an (equivariant) tensor train representation~\cite{oseledets_tensortrain_2011} of atomic energies with equal feature vectors $(v^1, \ldots, v^d) = v$.
An ETN potential in the tensor train format can then be written as follows
\begin{multline}\label{eq:etn_pot}
    V^{\rm ETN}({\bf \mathfrak{n}}_i)
    =
    \left(T^1_{(\ell_1' m_1' n_1') (\ell_1 m_1 n_1)} v_{(\ell_1' m_1' n_1')}({\bf \mathfrak{n}}_i)\right) \\
    \left(T^2_{(\ell_1 m_1 n_1) (\ell_2' m_2' n_2') (\ell_2 m_2 n_2)} v_{(\ell_2' m_2' n_2')}({\bf \mathfrak{n}}_i)\right) \\
    \ldots
    \left(T^d_{(\ell_{d-1} m_{d-1} n_{d-1}) (\ell_d' m_d' n_d')} v_{(\ell_d' m_d' n_d')}({\bf \mathfrak{n}}_i)\right)
\end{multline}
in which the channels $n_1,n_2,\ldots,n_d$ naturally emerge as ranks of the tensor network.
As feature vectors, we use the ones that contract the radial features before entering the tensor train~\cite{hodapp2024equivariant} as follows:
\[
    v_{(\ell m n)}
    =
    \sum_j \Big( B_{\ell n\alpha\lambda} Q_\alpha(\vert r_{ij} \vert) \big( A_{\ell\lambda\beta\gamma} z_\beta^i z_\gamma^j \big) \Big) Y_{\ell m}(r_{ij} / \vert r_{ij} \vert)
    .
\]

When comparing the moment tensor descriptors $M_{\mu,\nu}({\mathfrak{\bm n}}_i)$ and the feature vectors of ETNs, the radial parameters $c$ are decomposed into $A$ and $B$, and the angular part $r_{ij}^{\otimes \nu}$ is reshaped to $Y_{\ell m}(r_{ij} / \vert r_{ij} \vert)$. More details on the relation between MTP and ETN are provided in the original paper by Hodapp and Shapeev~\cite{hodapp2024equivariant}.

This enables learning similarities between the \emph{all} radial features from the data through the parameter tensors $A$ and $B$,
avoiding the problem of an exponentially growing size of the feature vectors~\cite{hodapp2024equivariant}.
As for MTP, the total energy of ETN is the sum of contributions $V^{\rm ETN}({\bf \mathfrak{n}}_i)$ for $N$ atoms.

The number of coefficients of the ETN potentials is proportional to $d \bar{r}^2 \bar{n}$, where $\bar{r}$ is the average rank of the tensor network, and $\bar{n}$ is the average size across all dimensions of the feature vector.
This makes the growth of their parameter space more controllable than, e.g., for MTPs.

\subsection{\label{sec:Fitting} Fitting}

For finding parameters of MLIPs, e.g. MTP or ETN parameters, we fit it on a (quantum-mechanical) training set. Let $K$ be a number of configurations in the training set and $N$ be a number of atoms. Denote a vector of MTP/ETN parameters to be found by ${\bm \theta}$. We find the optimal parameters ${\bm \bar{\theta}}$ by solving the following optimization problem (minimization of the objective function)
\begin{equation} \label{Fitting}
\begin{array}{c}
\displaystyle
\sum \limits_{k=1}^K \Bigl[ w_{\rm e} \left(E^{\rm MLIP}_k({\bm {\theta}}) - E^{\rm QM}_k \right)^2 +
\\
\displaystyle
w_{\rm f} \sum_{i=1}^{N} \left| {\bf f}^{\rm MLIP}_{i,k}({\bm {\theta}}) - {\bf f}^{\rm QM}_{i,k} \right|{^2} \Bigr]  \to \operatorname{min}.
\end{array}
\end{equation} 
We start from randomly initialized MLIP parameters. The optimal parameters ${\bm \bar{\theta}}$ are found numerically, using the iterative method to minimize the non-linear objective function, namely the Broyden-Fletcher-Goldfarb-Shanno algorithm. Thus, after optimization, the parameters ${\bm \bar{\theta}}$ are near the local minimum of the objective function. In this objective function, $E^{\rm QM}_k$ and ${\bf f}^{\rm QM}_{i,k}$ are the reference energies and forces, i.e., the ones to which we fit the MLIP energies $E^{\rm MLIP}_k$ and forces ${\bf f}^{\rm MLIP}_{i,k}$ and thus optimize the MLIP parameters ${\bm \theta}$. The factors $w_{\rm e}$ and $w_{\rm f}$ are non-negative weights which express the importance of energies and forces with respect to each other.
We took $w_{\rm e} = 1$ and $w_{\rm f} = 0.01$ to fit the ETN-based models and we took $w_{\rm e} = 1/N$ and $w_{\rm f} = 0.01$ to train the MTP-based models. These weights are default in the MLIP-2 and MLIP-4 codes, respectively. We refer to the minimization problem \eqref{Fitting} as MLIP fitting. 

\subsection{\label{sec:AL} Active Learning}
 
Assume that we found the vector of the optimal MTP parameters ${\bm \bar{\theta}}$ after solving \eqref{Fitting} and let the length of the vector be $m$. We then construct a matrix

\[
\mathsf{B}=\left(\begin{matrix}
\frac{\partial E^{\rm MTP}_1}{\partial \theta_1}({\bm {\bar{\theta}}}) & \ldots & \frac{\partial E^{\rm MTP}_1}{\partial \theta_m}({\bm {\bar{\theta}}}) \\
\vdots & & \vdots \\
\frac{\partial E^{\rm MTP}_K}{\partial \theta_1}({\bm {\bar{\theta}}}) & \ldots & \frac{\partial E^{\rm MTP}_K}{\partial \theta_m}({\bm {\bar{\theta}}}) 
\end{matrix}\right).
\]
From this matrix, we select a set of the $m$ most linearly independent rows, i.e., we find a submatrix $A$ with the maximum absolute value of the determinant (maximum volume). To find a submatrix $A$ of maximum volume from the matrix $B$, we use the MaxVol algorithm~\cite{goreinov2010-maxvol} which ensures the diversity of the configurations corresponding to the rows of $A$. 

After training the initial MTP and constructing the matrix $A$, we start a molecular dynamics (MD) simulation and compute the extrapolation grade for each configuration ${\bm x^*}$ created during this simulation
\begin{equation}
\label{eq:extrapol_grade}
    \gamma({\bm x^*}) = \max\limits_{1 \le j \le m} |c_j|,
\end{equation}
where the vector $\textbf{c} = (c_1, \ldots, c_m)$ for the configuration ${\bm x^*}$ is computed by
\begin{equation}
\label{eq:c}
    \textbf{c} = \Bigl(\frac{\partial E^{\rm MTP}}{\partial \theta_1} (\bm {\bar{\theta}}, {\bm x}^*), \dots, \frac{\partial E^{\rm MTP}}{\partial \theta_m} (\bm {\bar{\theta}}, {\bm x}^*) \Bigr) A^{-1}.
\end{equation}
We also introduce two thresholds: $\gamma_{\rm save}$ and $\gamma_{\rm break}$. Now we have all the prerequisites to formulate our AL algorithm.

\begin{enumerate}
    \item {\bf Run the MD simulation with extrapolation control.} The MD simulation algorithm generates configurations for which energy and forces are computed. An extrapolation grade is also calculated for each configuration. Depending on the degree of extrapolation $\gamma({\bm x^*})$ there are three possibilities: (i) if $\gamma({\bm x^*}) < \gamma_{\rm save} \approx 2$ then we proceed with the calculation of the energy and forces as we deal with interpolation or insignificant extrapolation; (ii) if $\gamma_{\rm save} < \gamma({\bm x^*}) < \gamma_{\rm break} \approx 10$ then we save the configuration to a file with the extrapolative configurations and proceed with the calculation of the energy and forces; (iii) if $\gamma({\bm x^*}) > \gamma_{\rm break}$ then we break the calculation as the extrapolation degree exceeds the critical threshold $\gamma_{\rm break}$.

    \item {\bf Selection of configurations.} In the previous step, the extrapolative configurations were saved to the file. However, the saved file may contain a large number of similar configurations. According to our active learning strategy, we select among the extrapolative configurations those that maximize the volume of the matrix $A$. Thus, in this step, we select the configurations to be added to the training set.
    
    \item {\bf Quantum-mechanical calculations.} After selecting the appropriate configurations, we calculate their quantum-mechanical energy and forces. 
    
    \item {\bf MTP retraining.} After \textit{ab intio} calculations of the configurations, we add them to the training set, retrain MTP, and update the matrix $A$.  
\end{enumerate}

We repeat steps 1-4 until no configuration is preselected during MD simulations. We note that the active learning algorithm can only be used with MTP and was implemented only in the MLIP-2 code~\cite{Novikov2021}. For the moment, we have not implemented a similar active learning algorithm for ETN.

\subsection{\label{sec:MTPs} Range separation}

In the present study, we added the D2~\cite{grimme2006} or D3~\cite{grimme2010} dispersion corrections to the local MTP and ETN potentials. We implemented the D2 correction in both the MLIP-2 and MLIP-4 codes, adding it to both MTP and ETN. The D3 correction was implemented only in the MLIP-4 code and combined only with ETN.

The local potentials, namely MTP and ETN, were trained on configurations with total energies and forces. In contrast, potentials that explicitly incorporate dispersion interactions via D2 or D3, namely, MTP+D2, ETN+D2, ETN+D3, were obtained in the following way. We first trained pure MTP and ETN models on configurations with D2 or D3 contributions subtracted from total energies and forces, and after training, dispersion correction was explicitly added to the top of a trained MLIP.

The explicit incorporation of dispersion via D2 or D3 allows to maintain a cutoff radius of 5--6~\AA\ for the machine learning part of these potentials, while the cutoff radius for D2 or D3 is set to 15~\AA\ for all simulations presented in this work.

\subsection{\label{sec:D2} D2 correction}

The D2 dispersion correction was introduced by Grimme et al.~\cite{grimme2006} to account for dispersion interactions that are not inherently included in the DFT framework.

D2 considers only the two-body dipole-dipole contribution to the dispersion energy, leading to the following form of energy expression:

\begin{equation}
E^{D2} = - s_6 \sum_{ij} \frac{C_6^{ij}}{r_{ij}^6} f_{\rm damp}(r_{ij}),
\label{d2_energy}
\end{equation}
where $i,j$ are indices of atoms, $s_6$ is a scaling factor dependent on a particular density functional, $C_6^{ij}$ is a coefficient describing interactions between atoms $i$ and $j$, $r_{ij}$ is an interatomic distance between atoms $i$ and $j$, and $f_{\rm damp}(r_{ij})$ is a damping function.

In this scheme, the coefficients $C_6^{ii}$ are precomputed parameters and are calculated from the empirical formula:
\begin{equation}
    C_6^{ii} = 0.05NI_A\alpha_A^0,
\end{equation}
where $N$ is a scaling factor, $I_A$ is the first ionization potential calculated with PBE0, and $\alpha_A^0$ is static dipole polarizability. The $C_6^{ij}$ are defined as the geometric mean of $C_6^{ii}$ and $C_6^{jj}$:

\begin{equation}
    C_6^{ij} = \sqrt{C_6^{ii}C_6^{jj}}.
\end{equation}

The Fermi damping function $f^{\rm Fermi}_{\rm damp}$ is utilized in D2:
\begin{equation}
    f^{\rm Fermi}_{\rm damp} = \frac{1}{1+e^{{20(r_{ij}/(s_R R_{\rm vdW})-1)}}},
\end{equation}
where $s_R$ is a scaling parameter, $R_{\rm vdW}$ is a sum of van der Waals radii of interacting atoms.

However, in our implementation of the D2 correction (as well as the D3 correction), we chose Becke-Johnson damping function $f^{(n)}_{\rm damp, BJ}$:
\begin{equation}
    f^{(n)}_{\rm damp, BJ} = \frac{r^n}{r^n+(a_1R_0+a_2)^n}
\label{bj_damping}
\end{equation}
with $n=6$. We adjusted the coefficients $a_1$, $a_2$, and $R_0$ on CCl$_4$, CH$_4$, and toluene dimer binding curves.

\subsection{\label{sec:D3} D3 correction}

In contrast to the D2 scheme, the D3 dispersion correction~\cite{grimme2010} explicitly takes into account information about the atomic environment via fractional coordination number. Moreover, in the D3 approach, the dispersion energy includes not only the dipole-dipole interaction term, proportional to $1/r^6$, but also the higher order dipole-quadrupole contribution proportional to $1/r^8$:

\begin{equation}
E^{D3} = - \sum_{ij}\sum_{n=6,8} s_n \frac{C_n^{ij}}{r_{ij}^n} f_{\rm damp}^{(n)}(r_{ij}),    
\label{d3_energy}
\end{equation}
where $i,j$ are the indices of atoms, $s_n$ is a scaling factor dependent on the density functional, $C_n^{ij}$ is a coefficient describing interactions between atoms $i$ and $j$, $r_{ij}$ is the interatomic distance between atoms $i$ and $j$, and $f_{\rm damp}^{(n)}(r_{ij})$ is damping function, dependent on the power $n$.

It is possible to include the three-body interaction term in the expression of the D3 energy. However, in our implementation, we limited D3 to the two-body form, as this implementation is widely used and the contribution of the three-body interaction term is typically much lower than the two-body term~\cite{grimme2016}.

In the D3 correction, $C_n^{ij}$ explicitly includes information on the environments of both $i$ and $j$ atoms via coordination numbers:
\begin{equation}
CN^i = \sum_{i \neq j}^{N_{\rm atoms}}\frac{1}{1+ e^{{-16(4(R_{i,{\rm cov}}+R_{j,{\rm cov}})/(3 r_{ij})-1)}}},
\end{equation}
where $R_{i,{\rm cov}}$, $R_{j,{\rm cov}}$ are covalent radii of elements~\cite{pyykko2009molecular}.

The computed coordination numbers are further used for interpolating $C_6^{ij}$ based on a set of precomputed coefficients $C_{6, \rm ref}^{ij}$, obtained for reference compounds~\cite{grimme2010}:
\begin{equation}
    C_6^{ij}(CN^i, CN^j) = \frac{\sum_p^{N_i}\sum_q^{N_j} C_{6, \rm ref}^{ij}(CN_p^i, CN_q^j)L_{pq}}{\sum_p^{N_i}\sum_q^{N_j}L_{pq}},
\end{equation}
where $N_i$, $N_j$ are numbers of reference compounds for elements $i$ and $j$, and $L_{pq}$ is defined as:
\begin{equation}
    L_{pq} = e^{{-4\big( (CN^i - CN^i_p)^2 + (CN^j - CN^j_q)^2 \big)}}.
\end{equation}

Based on the obtained $C_6^{ij}$, the coefficients $C_8^{ij}$ for the higher-order term in dispersion energy \eqref{d3_energy} is computed:
\begin{equation}
    C_8^{ij} = 3C_6^{ij}\sqrt{Q_iQ_j},
\end{equation}
where $Q_i$, $Q_j$ are precomputed parameters derived from atomic densities and atomic numbers of elements $i$ and $j$~\cite{grimme2010}.

For the D3 correction, two types of damping function can be utilized: zero-damping and Becke-Johnson damping (BJ)  \eqref{bj_damping}. For our implementation of D3 we chose the BJ damping with $n=6$ for the $1/r^6$ damping term, and $n=8$ for the higher order term.

\subsection{Protocol of datasets generation and training of MLIPs}
\begin{figure}[!ht]
    \centering
    \includegraphics[width=1\linewidth]{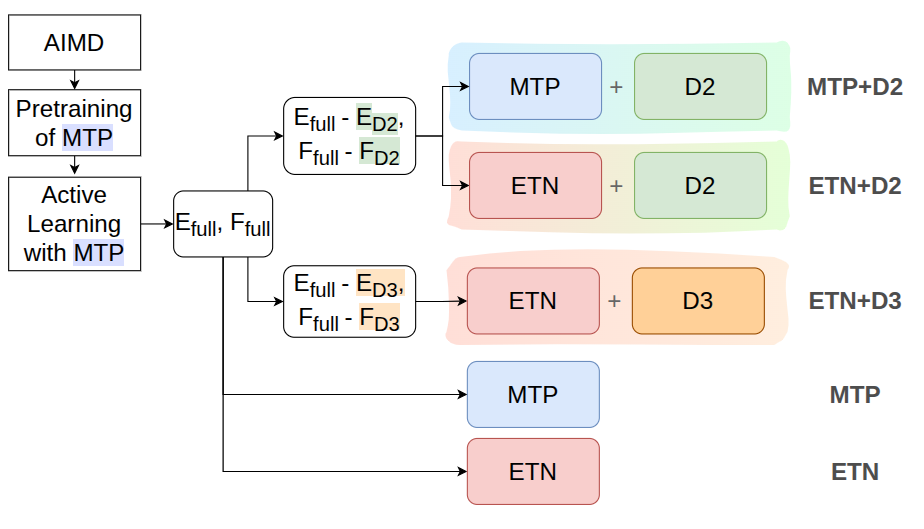}
    \caption{The general scheme for generating training sets for all the MLIPs considered. At the first step, the AIMD simulation is conducted and initial MTP is trained on the configurations uniformly chosen from the AIMD trajectory. Next, a pretraining set formed from uniformly selected configurations and configurations, additionally selected from the AIMD trajectories with the MaxVol algorithm~\cite{Podryabinkin2016}. The initial MTP retrained on the resulting pretraining set. After that, the AL of MTP during MD simulations is carried out, and results in the actively selected training set with total energies and forces. Further steps correspond to training of MTP and ETN on the dataset including total energies and forces, and subtracting D2 or D3 contributions and utilizing the resulting training sets for fitting of MTP or ETN part of MTP+D2, ETN+D2, ETN+D3 potentials.}
    \label{fig:datasets_generateion}
\end{figure}

We collected training sets with total energies and forces via the AL approach described above and implemented in the MLIP-2 code for MTP.
For fitting of models that explicitly incorporate dispersion via D2 or D3 (MTP+D2, ETN+D2, ETN+D3), the training set with total energies and forces was post-processed by subtracting D2 or D3 contributions from the total energies and forces. The general workflow for obtaining training sets for each of the considered models is shown in Figure \ref{fig:datasets_generateion}.

We started creating a training set from \textit{ab initio} molecular dynamics (AIMD) simulations. An initial configuration was generated using Packmol~\cite{martinez2009packmol}. Next, we uniformly chose about 20 configurations from this AIMD trajectory to avoid correlations between configurations and fit an initial MTP. After that, we selected configurations using the MaxVol algorithm from the pool of configurations generated during AIMD, updated the initial training set, and refitted the MTP, obtaining pretrained MTP. Subsequently, we actively trained the pretrained MTP during 10 parallel MD simulations (AL-MD) of 100~ps each using the AL procedure described above, utilizing LAMMPS package~\cite{LAMMPS} for MD. As a result, we obtained the actively selected training set with total energies and forces. During both AIMD and MD in the AL procedure, a timestep was set to 1~fs.

The collected training set was then used to train the MTP and ETN models. After subtracting the contributions of D2 or D3 from the total energies and forces in the collected training set, we obtained training sets for the MTP+D2, ETN+D2, and ETN+D3 models.

For all the investigated systems, the level of the MTP-based potentials (MTP and MTP+D2) was set to 16, corresponding to 222 parameters to be optimized during the fitting, and for the ETN-based models (ETN, ETN+D2, and ETN+D3) we took 124 coefficients to be fitted. We note that the AL procedure was performed only for MTP and the resulting training set may not be optimal for other tested MLIPs, especially for the ETN-based models. We emphasize that in this procedure we implicitly assume, since ETN has fewer parameters, that the training set actively selected by MTP would also be suitable for ETN.

For creating validation sets, the configurations were uniformly sampled from independent 100~ps MD simulation conducted with actively trained MTP. The energies and forces for the sampled configurations were then computed using \textit{ab initio} calculations. The ratio between the size of the training and validation sets was set to 8:2.

\subsection{Assessing the quality of MLIPs}

To compare the quality of the fitted MLIPs we evaluated training and validation errors of predicting energies and forces.
In addition, we designed two series of tests. In the first series, we evaluated the accuracy of potentials in predicting intermolecular interactions. To that end, we predicted binding curves for a diverse set of dimers and compared them with the ones obtained with \textit{ab initio} calculations. We calculated the deviation between the position and depth of the minima, as well as the full width of the well at half its depth (referred to as full-width at half-minimum or FWHM) obtained with potentials and \textit{ab initio} calculations (see supplementary material for details). In this test, the binding curves were represented as dependencies of the interaction energy on the distance between the centers of mass of individual molecules. The second series of tests aimed to assess the accuracy of density and RDF calculated during MD, performed with LAMMPS package~\cite{LAMMPS}. We used experimental data as reference.

For evaluating the errors, we used the root mean square error (RMSE) and mean absolute percentage error (MAPE):

\begin{equation}
    {\rm RMSE} = \sqrt{\frac{\sum_{i=1}^N (x_i-\hat x_i)^2}{N}},
\end{equation}

\begin{equation}
    {\rm MAPE} = \frac{\sum_{i=1}^N |x_i-\hat x_i|}{N},
\end{equation}
where $N$ is the number of samples, and $x_i$ and $\hat x_i$ are the predicted and true values, respectively.

To ensure that error estimations are statistically meaningful, we conducted each test for an ensemble of five potentials for each type of MLIP model considered, except for calculating density and RDF with ETN+D3, as this type of MLIP is more computationally demanding than others.

\section{Results and discussion}

In this section, we present results of application of the described MLIPs to the modeling of liquid carbon tetrachloride, liquid methane, and liquid toluene. For all systems, we focus on evaluating the accuracy differences between MLIP(l)'s and MLIP+D's. 

We note that for carbon tetrachloride, we performed two test cases. In the first case, we compared the accuracy of MLIP(l)'s and MLIP+D's when the $R_{\rm cut}$ for all potentials is set to 5.5~\AA, which limits the ability of MLIP(l)'s to fully capture the range of dispersion interactions. In the second case, we explored to what extent the accuracy of MLIP(l)'s can be improved by increasing the $R_{\rm cut}$ to 7.5~\AA, allowing them to capture the major part of long-range interactions (while maintaining a cutoff radius of 5--6~\AA\ for MLIP+D's). For methane and toluene, we took only the $R_{\rm cut}$ of 7.5~\AA\ for the MLIP(l)'s and of 6~\AA\ for MLIP+D's.

\subsection{Computational details}\label{comptational_details}

For the generation of the training set, we employed the AL scheme described above (Figure \ref{fig:datasets_generateion}). The simulation setup for AIMD and AL-MD, details of the \textit{ab initio} calculations, and sizes of the obtained training sets are presented in Table \ref{tab:dataset_generation}.

\begin{table*}[!ht]
    \begin{center}
    \begin{tabular}{p{1.1cm}|p{1.5cm}|p{1.1cm}|p{2.3cm}|p{1.9cm}|p{1.2cm}|p{1.2cm}|p{1.2cm}|p{3.5cm}|p{1.5cm}}
        \hline
        \hline
         & \textit{ab initio} package& level of theory& basis set & pseudo-potential & cutoff, \newline k-points & number of\newline  molecules\newline in cell& cell size, \AA & MD& training \newline set size\\ \hline
         CCl$_4^{(a)}$ & CP2K~\cite{kuhne2020cp2k}&PBE-D3 & Gaussian Plane Waves~\cite{lippert1997hybrid},\newline QZV2P-GTH-q4,\newline QZV2P-GTH-q7& GTH-PBE-q4,\newline GTH-PBE-q7& energy cutoff~\cite{vandevondele2005quickstep}: 400~Ry; $\Gamma$-point& 6 & 10-12   & AIMD: at 298~K for 1~ps in the NVT ensemble;\newline
AL-MD: NpT ensemble at 293~K and 1~bar& 478\\ \hline
         CCl$_4^{(b)}$ & VASP~\cite{kresse1996software}& PBE-D3 & Plane Waves (PW) & PAW\_PBE  & 900 eV, \newline $\Gamma$-point & 11 & 15-17 & AIMD: 500~ps was carried out in the NVE, velocities initialized at 298~K;\newline
AL-MD: NpT ensemble at 293~K and 1~bar& 850 \\ \hline
         CH$_4$ & VASP & PBE-D3  & PW & PAW\_PBE &900 eV\newline $\Gamma$-point & 57 & 15-18 & AIMD: NVE ensemble for 500~fs, velocities initialized at 298~K;\newline
AL-MD: NpT ensemble at 150~K and 10~bar.& 306 \\ \hline
         Toluene & FHI-aims~\cite{blum2009ab}& PBE-TS & Numerically tabulated Atom-centered Orbitals~\cite{blum2009ab},  intermediate basis set& all electrons & 2x2x2 k-point grid& 13 & 15-18 & AIMD: NVT ensemble for 400~fs, at 298~K;\newline
AL-MD: NpT ensemble at 298~K and 1~atm& 520 \\
        \hline 
        \hline
        \end{tabular}
    \end{center}
    \caption{Computational details on the generation of the training sets.}
    \label{tab:dataset_generation}
\end{table*}
In all \textit{ab initio} calculations, we employed dispersion-corrected DFT at the PBE-D3 or PBE-TS level of theory. The employed D3 version includes the Becke-Johnson damping function and omits the three-body (ATM) term. This means that we do not account for many-body effects, which affect the accuracy of the ab initio method and, consequently, the trained potentials. However, since dispersion in ab initio methods is described only by two-body potentials, this approach is consistent with our use of two-body D2 and D3 terms in the MLIP+D framework. However, if MLIP+Dn models were fitted to reference data that included many-body effects, incorporating the ATM term in the D3 term should be expected to enhance the accuracy.

The chosen \textit{ab initio} models described above were also employed for the calculation of the dimer binding curves, used to test the resulting MLIPs. In addition to testing MLIPs on the binding curves, we also used them during MD simulations to predict densities and RDFs. To determine the densities, we equilibrated the simulation cells in the NpT ensemble until convergence was achieved, and then averaged the density over 100~ps of the simulation. The simulation details of the density calculations are given in Table~\ref{tab:comp_details_density}. For calculating RDFs, the simulations started from the configuration with converged density. Then, the system was equilibrated in the NVT ensemble at the target temperature during 10~ps. Subsequently, the ensemble was changed to NVE, and the RDF was averaged over 100~ps. A timestep of 1~fs was used in all simulations.
For CCl$_4$ and toluene, the RDF was obtained under the same conditions as the density, while for CH$_4$ the temperature was set to 92~K and the pressure was set to 0.13~atm to match the conditions of the available experimental data.

\begin{table}[!ht]
    \begin{center}
    \begin{tabular}{c|c|c|c|c}
        \hline
        \hline
        & \multirow{2}{1.5cm}{number of molecules} & \multicolumn{3}{c}{Density simulation} \\ \cline{3-5}
        &   & trajectory length, ps & T, K & p, atm \\ \hline
        CCl$_4^{(a)}$ & 162 & \multirow{2}{*}{500} & \multirow{2}{*}{293} & \multirow{2}{*}{1}  \\ \cline{1-2}
        CCl$_4^{(b)}$ & 168 &  &  &  \\ \hline
        CH$_4$ & 456 & 300 & 110 & 20.48  \\ \hline
        Toluene & 104 & 200 & 298 & 1 \\ \hline
        \hline 
        \end{tabular}
    \end{center}
    \caption{Computational details on density calculations.}
    \label{tab:comp_details_density}
\end{table}

\subsection{Training and validation errors}

\begin{figure*}[!ht]
    \centering
    \includegraphics[width=1\linewidth]{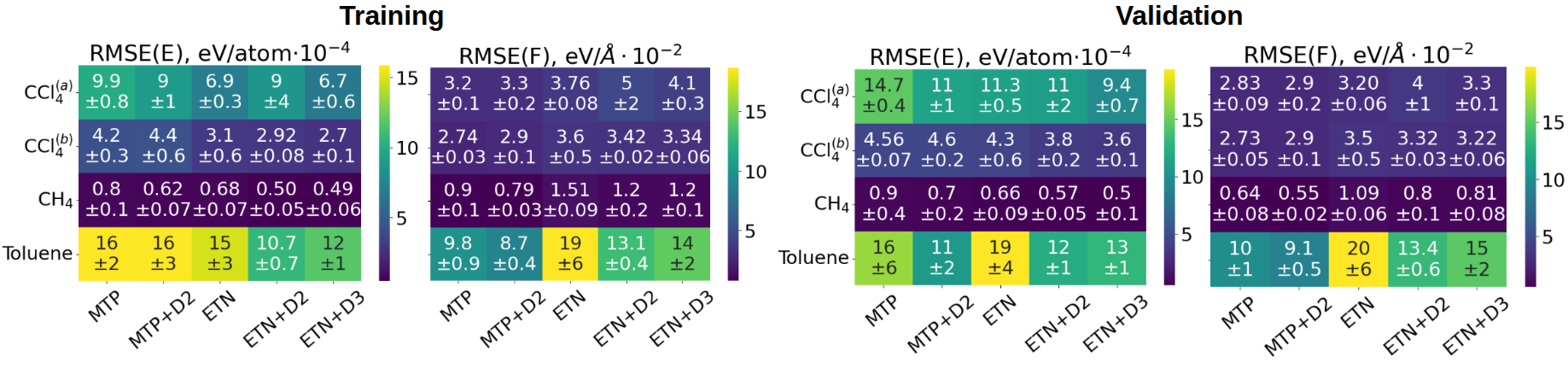}
    \caption{Training and validation RMSEs with one-$\sigma$ confidence interval, calculated for: CCl$_4$ using MLIPs with $R_{\rm cut}=5.5$~\AA\ (CCl$_4^{(a)}$), CCl$_4$ using MLIP(l)'s with $R_{\rm cut}=7.5$~\AA\ and MLIP+D's with $R_{\rm cut}=6$~\AA\ (CCl$_4^{(b)}$), CH$_4$ using MLIP(l)'s with $R_{\rm cut}=7.5$~\AA\ and MLIP+D's with $R_{\rm cut}=6$~\AA, and toluene using MLIP(l)'s with $R_{\rm cut}=7.5$~\AA\ and MLIP+D's with $R_{\rm cut}=6$~\AA.}
    \label{fig:train_val_errors}
\end{figure*}

For all the test cases, ensembles of five potentials were trained for each MLIP type. The training and validation RMSEs, averaged over the ensembles, are presented in Figure \ref{fig:train_val_errors}. 
For MLIPs with a cutoff of 5.5~\AA, we observe a reduction in the energy RMSEs moving from MTP to MTP+D2, while the effect of dispersion correction for ETN-based potentials is less significant and observed only for validation energy RMSEs. 
When extending the cutoff radius of MLIP(l)'s from 5.5~\AA\ to 7.5~\AA, the errors decreased by approximately a factor of three for energy, matching the accuracy of MLIP+D's.
This trend also holds for CH$_4$, where errors are approximately four times smaller for energies and two times lower for forces than those obtained for CCl$_4$ with MLIPs with extended $R_{\rm cut}$. 
However, the MLIPs fitted on toluene, even when the cutoff radius of MLIP(l)'s is extended, were not able to reach the accuracy level of MLIP+D's.

\subsection{Binding curves}

\begin{figure*}[!ht]
    \centering
    \includegraphics[width=1\linewidth]{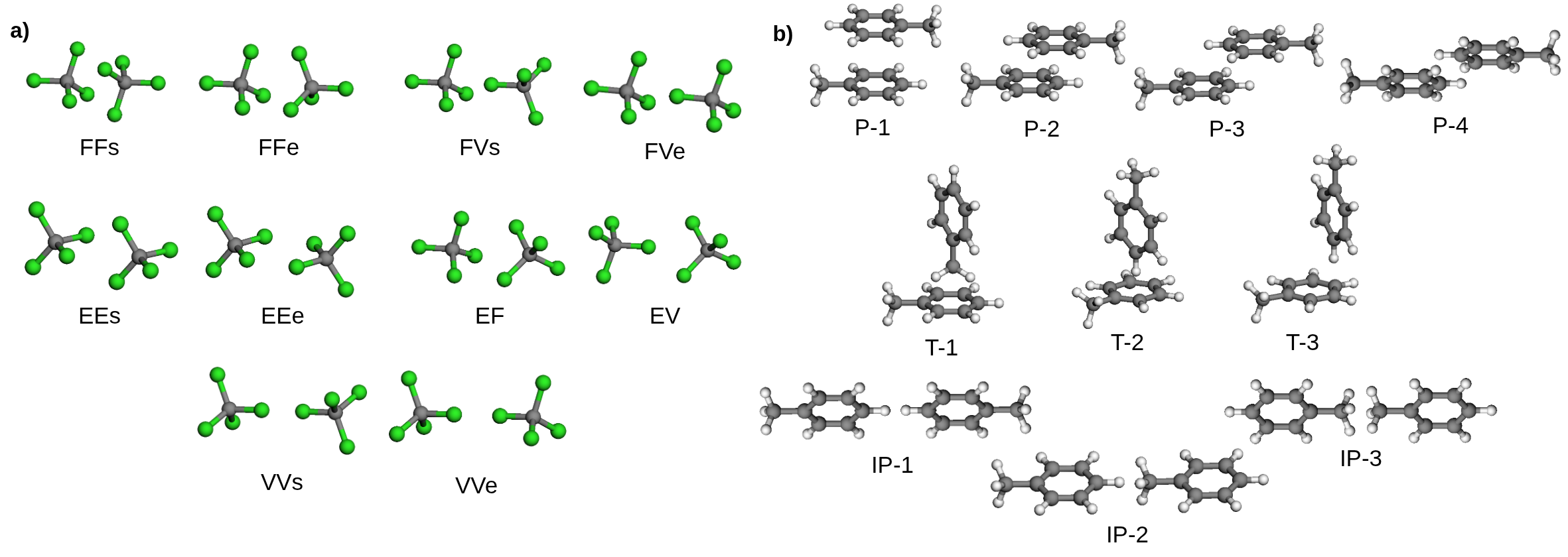}
    \caption{a) Dimers of CCl$_4$ and CH$_4$ used for generation of the binding curves. The labels correspond to description of mutual orientation, based on treatment of CCl$_4$ molecule as tetrahedron. F corresponds to ``face'', V -- ``vertex'', E -- ``edge'', and, finally, s and e correspond to staggered and eclipsed conformations; b) dimers of toluene used for constructing binding curves and minima depths and positions. The labels denote mutual orientations of molecules: "P" -- parallel, "T" -- T-shaped, "IP" -- in plane.}
    \label{fig:dimers}
\end{figure*}

To assess how the observed RMSEs relate to the accuracy in predicting the potential energy surface, we benchmarked the potentials on the binding curves prediction. To this end, we considered diverse sets of dimers (Figure \ref{fig:dimers}) to evaluate the precision across a range of interaction strengths and distances (Table \ref{tab:dimers}). For CCl$_4$ and CH$_4$, we constructed the dimer set by enumerating configurations with varying numbers of Cl$\dotsi$Cl or H$\dotsi$H contacts (Figure \ref{fig:dimers} (a)). For toluene, the dimers included configurations with and without stacking interactions, T-shaped dimers, and dimers with molecules lying in the same plane (Figure \ref{fig:dimers} (b)).

The minima depth and position on the binding curves for CCl$_4$, CH$_4$ and toluene dimers obtained with {\it ab initio} calculations are presented in Table \ref{tab:dimers}. The binding curves were calculated with the same \textit{ab initio }package that was used to generate the corresponding training set. Compared to carbon tetrachloride, CH$_4$ dimers exhibit closer and shallower minima. In contrast, toluene dimers possess the widest range of minima depths and positions among the studied systems, suggesting that toluene will be the most challenging molecule to model for MLIPs.

\begin{table}[!ht]
    \centering
    \begin{tabular}{c|c|c|c|c|c|c|c|c|c|c|c}
        \hline
        \hline
         & & FFs  & FFe & FVs & FVe & EEs & EEe & EF & EV & VVs & VVe \\ \hline
        \multirow{2}{1.0cm}{CCl$_4^{(a)}$} & E& -95 & -74 & -66 & -66 & -44 & -72 & -73 & -62 & -21 & -18 \\ \cline{2-12}
        & r & 4.9 & 5.2 & 5.9 & 5.9 & 6.0 & 5.5 & 5.4 & 6.2 & 7.3 & 7.3   \\ \hline \hline
        \multirow{2}{1.0cm}{CCl$_4^{(b)}$} & E& -93 & -75 & -63 & -62 & -48 & -69 & -71 & -59 & -20 & -21 \\ \cline{2-12}
        & r& 4.9 & 5.2 & 5.9 & 5.9 & 6.0 & 5.5 & 5.4 & 6.2 & 7.2 & 7.2   \\
         \hline
         \hline
        \multirow{2}{1.0cm}{CH$_4$} & E & -35 & -33 & -29 & -29 & -25 & -29 & -31 & -25 & -17 & -18 \\ \cline{2-12}
         & r & 3.7 & 3.8 & 4.1 & 4.1 & 4.0 & 4.0 & 3.9 & 4.2 & 4.6 & 4.6   \\
         \hline
         \hline
        \multirow{3}{1.0cm}{Toluene} & & P-1  & P-2 & P-3 & P-4 & T-1 & T-2 & T-3 & IP-1 & IP-2 & IP-3 \\ \cline{2-12}
        & E & -173 & -172 & -134 & -115 & -123 & -139 & -104 & -26 & -46 & -57 \\ \cline{2-12}
        & r & 3.8 & 4.2 & 5.1 & 5.8 & 6.0 & 5.6 & 5.9 & 8.2 & 8.3 & 8.4   \\ 
        \hline 
        \hline
        \end{tabular}
    \caption{Depths and positions of binding curves minima for carbon tetrachloride, methane, and toluene dimers. For carbon tetrachloride binding curves were calculated with both CP2K (CCl$_4^{(a)}$) and VASP (CCl$_4^{(b)}$) accordingly to the \textit{ab initio} package used for the datasets generation (Table \ref{tab:dataset_generation}). Minimum energies (E) are presented in meV, while minimum positions (r) in \AA}
    \label{tab:dimers}
\end{table}

\begin{figure*}[!ht]
    \centering
    \includegraphics[width=1\linewidth]{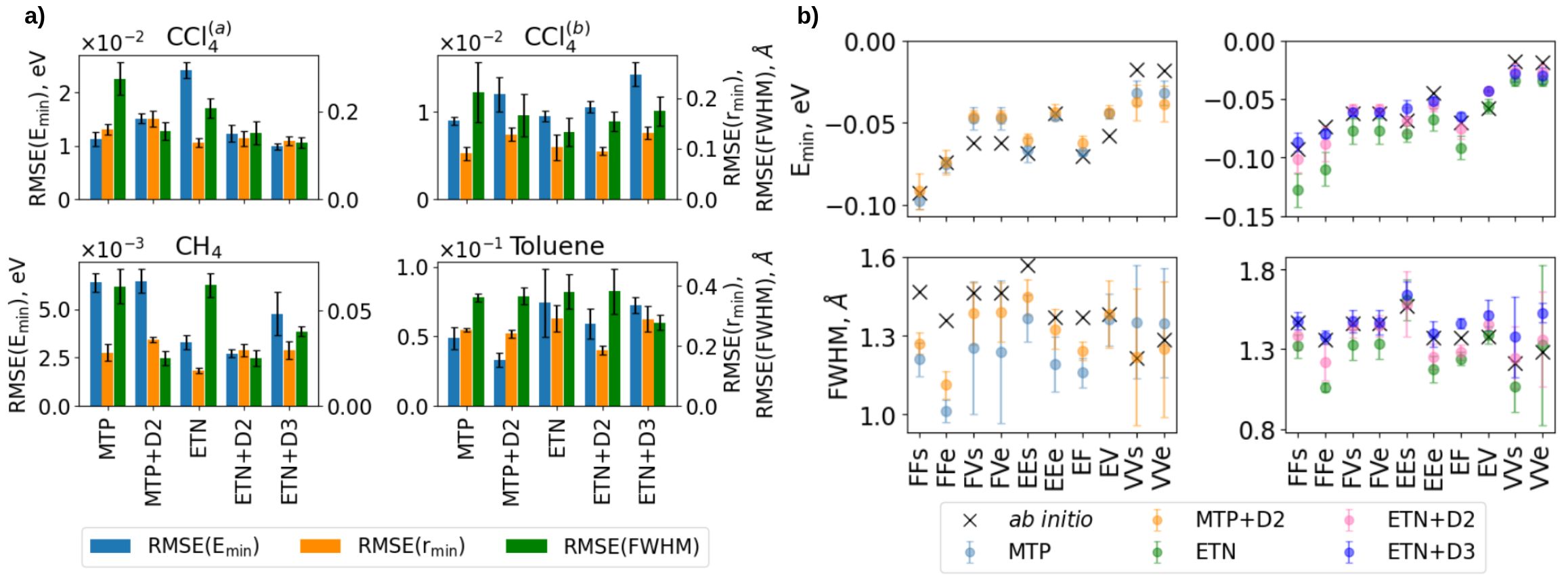}
    \caption{a) RMSEs evaluated for minima energy (E$_{\rm min}$), position (r$_{\rm min}$) and FWHM for all the considered test cases, involving modeling of: CCl$_4$ using MLIPs with $R_{\rm cut}=5.5$~\AA\ (CCl$_4^{(a)}$), CCl$_4$ using MLIP(l)'s with $R_{\rm cut}=7.5$~\AA\ and using MLIP+D's with $R_{\rm cut}=6$~\AA\ (CCl$_4^{(b)}$), CH$_4$ using MLIP(l)'s with $R_{\rm cut}=7.5$~\AA\ and using MLIP+D's with $R_{\rm cut}=6$~\AA, and toluene using MLIP(l)'s with $R_{\rm cut}=7.5$~\AA\ and using MLIP+D's with $R_{\rm cut}=6$~\AA; b) absolute values E$_{\rm min}$ and FWHM, calculated with \textit{ab initio} method and predicted for CCl$_4$using MLIPs with $R_{\rm cut}=5.5$~\AA ~(CCl$_4^{(a)}$).}
    \label{fig:curves_errors}
\end{figure*}

The calculated RMSEs for the energy minimum ($E_{\rm min}$), the position of its minimum ($r_{\rm min}$), and the FWHM are shown in Figure~\ref{fig:curves_errors}~(a). We observed that explicit incorporation of the dispersion correction significantly affects the FWHM RMSEs when $R_{\rm cut}$ is set to 5.5~\AA\ for MLIP(l)'s, and it also significantly improves the accuracy of $E_{\rm min}$ for the ETN model (Figure~\ref{fig:curves_errors}~(a)). The effect of incorporating the dispersion correction corresponds to approximately 50\% decrease in the FWHM RMSE for MTPs-based potentials and to 50\% decrease in the energy RMSE for the ETN-based potentials. However, RMSE changes of the same order of magnitude are observed when extending the MLIP(l)'s cutoff radius to 7.5~\AA\ (Figure~\ref{fig:curves_errors}~(a)). In particular, the ETN energy RMSE and MTP FWHM RMSE decrease by about 50\%, reaching the accuracy of MLIP+D's. Accordingly, for the other systems, since the MLIP(l)'s cutoff radii were extended, the difference between MLIP(l)'s and MLIP+D's is not significant in any property excluding FWHM in CH$_4$ where MLIP+D's still show higher accuracy than MLIP(l)'s. 
At the same time, the RMSEs for CH$_4$ are much smaller than for CCl$_4$, while for toluene they are significantly higher than in other test cases, including CCl$_4$ modeling with MLIPs when $R_{\rm{cut}}$ is set to 5.5~\AA.
Additionally, we note that predicting the depth and FWHM is more challenging than predicting the position of the minimum, as it is less sensitive to the incorporation of dispersion correction or increasing $R_{\rm cut}$. 

Since we observed the significant impact of dispersion correction in modeling of CCl$_4$ with MLIPs when $R_{\rm cut}$ is set to 5.5~\AA, we quantified this impact in absolute values of the energy minima and FWHMs (see supplementary material for absolute values of the minima energy and FWHM for other test cases). As shown in Figure \ref{fig:curves_errors}~(b), the high energy RMSE of ETN corresponds to a significant overbinding of the majority of dimers, which was mitigated by the explicit incorporation of the dispersion correction in ETN+D2 and ETN+D3. At the same time, the difference between the energies predicted by MTP and MTP+D2 is not significant. We also note that all the potentials, excluding ETN+D3, tend to underestimate FWHM, especially the MTP and ETN potentials.

\begin{figure}[!ht]
    \centering
    \includegraphics[width=1\linewidth]{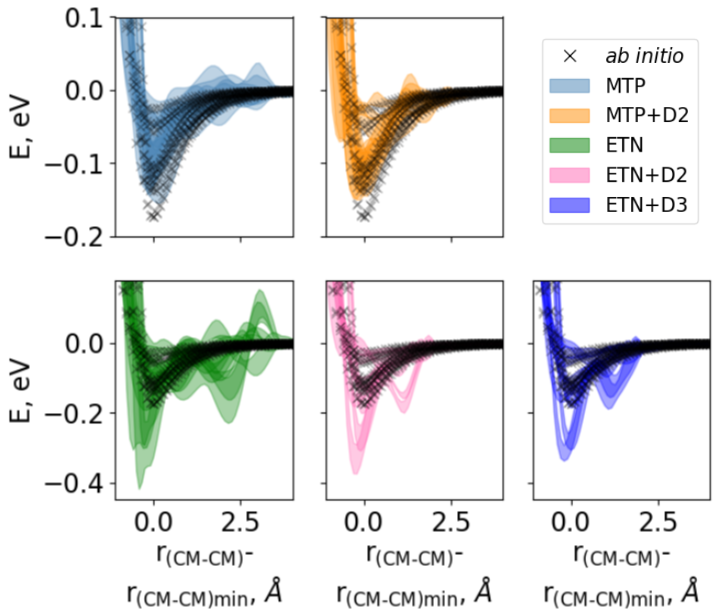}
    \caption{Toluene dimers binding curves predicted by MLIPs compared to \textit{ab initio} calculated. }
    \label{fig:tol_curves}
\end{figure}

In addition to the findings related to the energy minima and the potential well width, we made a further observation concerning the energy dependence on the intermolecular separation.
In Figure \ref{fig:tol_curves} we demonstrate the unphysical oscillations in the toluene dimer binding curves predicted by MLIP(l)'s with $R_{\rm cut}=7.5$~\AA\ and MLIP+D's with $R_{\rm cut}=6$~\AA. This behavior is observed for all the MLIPs, but is more severe in the case of the ETN-based ones. These oscillations are weaker for MLIP+D's compared to MLIP(l)'s. This observation explains the difference in MLIP(l)'s and MLIP+D's training and validation energy errors, which could not be explained from the energy minima, position, and FHWM RMSEs, as they did not change substantially with the incorporation of dispersion correction.
Similarly, such oscillations are observed for CCl$_4$ VVe dimer binding curve predicted by MTP with the extended $R_{\rm cut}$ (see supplementary material for details), while MTP+D2 preserved more correct asymptotics. We note that all binding curves for CCl$_4$ and CH$_4$ are also given in supplementary material.

Finally, the overall improvement in binding curve predictions achieved by incorporation of long-range interactions is both intuitive and well-documented in literature for MLIPs that account for electrostatic interactions~\cite{grisafi2021multi,yue2021short}.

\subsection{Densities}

Next, we calculated the liquid densities for all considered cases. Similar to the binding curve tests, we focus on the impact of adding the dispersion correction and, for CCl$_4$, on extending the cutoff radius for MLIP(l)'s.

\begin{figure*}[!ht]
    \centering
    \includegraphics[width=1\linewidth]{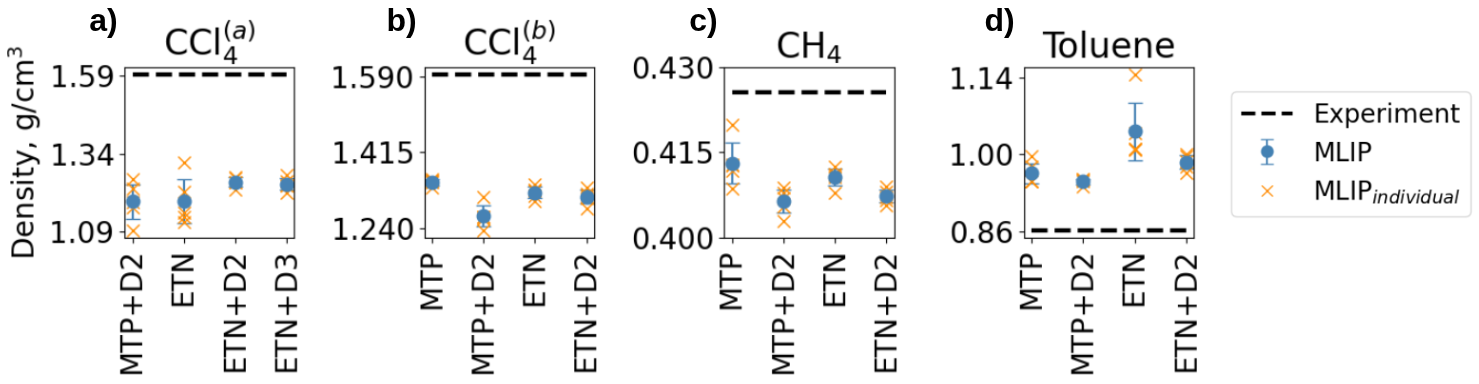}
    \caption{Densities obtained from experiment~\cite{hien1978density, harris1980density, harris2000temperature} and mean densities, their 1-$\sigma$ confidence intervals, obtained with ensembles of MLIPs, along with predictions of individual MLIPs for: a) CCl$_4$ using MLIPs with cutoff 5.5~\AA\ (CCl$_4^{(a)}$); b) CCl$_4$ using MLIP(l)'s with cutoff 7.5~\AA\ and MLIP+D's with cutoff 6~\AA\_re (CCl$_4^{(b)}$); c) CH$_4$; d) toluene.}
    \label{fig:densities}
\end{figure*}

All density values were obtained by averaging over 100~ps of the NpT MD trajectories after the density converged. The details of the simulation are given in Table~\ref{tab:comp_details_density}. Figure~\ref{fig:densities} shows the predicted and experimental densities, and Table~\ref{tab:densities} contains the corresponding MAPEs for all investigated systems. Similarly to the binding curve tests, only CCl$_4$ was modeled with MLIP(l)'s and MLIP+D's using two sets of cutoff radii.

\begin{table}[!ht]
    \centering
    \begin{tabular}{c|c|c|c|c|c}
        \hline
        \hline
         & MTP  & MTP+D2 & ETN & ETN+D2 & ETN+D3\\ \hline
         \multirow{2}{1.2cm}{CCl$_4$, (a)}& 0.991 & 0.26& 0.26& 0.216 & 0.22\\
         & $\pm$ 0.001 &  $\pm$ 0.04&  $\pm$ 0.04&  $\pm$ 0.009 &  $\pm$ 0.02\\ \hline
         \multirow{2}{1.2cm}{CCl$_4$, (b)} & 0.156& 0.20& 0.170 & 0.18& \multirow{2}{*}{-} \\ 
         & $\pm$ 0.004&  $\pm$ 0.02&  $\pm$0.008 & $\pm$ 0.01&  \\  \hline
         \multirow{2}{1.2cm}{CH$_4$} & 0.033  & 0.045  & 0.035  & 0.043 & \multirow{2}{*}{-} \\ 
         & $\pm$ 0.004 &  $\pm$ 0.005 &  $\pm$ 0.003 &  $\pm$ 0.002 & \\ \hline
         \multirow{2}{1.5cm}{Toluene} & 0.12& 0.100& 0.21  & 0.14& \multirow{2}{*}{-} \\ 
         & $\pm$ 0.02&  $\pm$ 0.006 &  $\pm$ 0.06&  $\pm$ 0.01& \\ \hline
         \hline
    \end{tabular}
    \caption{Density MAPE and its one-$\sigma$ confidence interval, compared to experiment for CCl$_4$~\cite{hien1978density}, CH$_4$~\cite{harris1980density}, and toluene~\cite{harris2000temperature}. Here CCl$_4^{(a)}$ corresponds to the results obtained using MLIPs with cutoff 5.5~\AA, while CCl$_4^{(b)}$ refers to results obtained using MLIP(l)'s with cutoff 7.5~\AA\ and MLIP+D's with cutoff 6~\AA.}
    \label{tab:densities}
\end{table}

The two key findings during the modeling of the CCl$_4$ system using all MLIPs with cutoff 5.5~\AA are given below. First, the MTP potential does not maintain stable molecular dynamics in the liquid phase, as we observed the density collapsing to near zero values. However, this issue is resolved by explicit incorporation of dispersion interactions in MTP+D2, resulting in a 25.6\% deviation from the experimentally measured density~\cite{hien1978density} of 1.594 g/cm$^3$ (Figure \ref{fig:densities} (a), Table \ref{tab:densities}). Unlike MTP, the ETN potential is capable of preserving the liquid phase, likely due to the overbinding of molecules, which was demonstrated in the dimer binding curves test (Figure \ref{fig:curves_errors}~(b)). Additionally, ETN+D2 and ETN+D3 have a higher accuracy compared to ETN, achieving a deviation of approximately 22\% from the experimental density. Secondly, unlike the results of the binding curves tests, ETN+D3 does not outperform ETN+D2 in predicting density.

When the cutoff radii were extended, we obtained a substantial improvement over the MLIPs with smaller cutoff radii, as can be seen from the density values and MAPEs (Figure \ref{fig:densities} (b), Table \ref{tab:densities}). Furthermore, similarly to the binding curves, the errors of all MLIPs are close to each other. Thus, we demonstrated that it is sufficient to introduce an $R_{\rm cut}$ capturing the major part of long-range interactions for MLIP(l)'s for accurate modeling of liquid carbon tetrachloride, since the difference in accuracy between the MLIP(l) and MLIP+D potentials is minor for all properties calculated.

In view of all the results for CCl$_4$, for the other two systems, we decided to limit the tests for MLIP+D's with $R_{\rm cut}=6$~\AA\ and MLIP(l)'s with $R_{\rm cut}=7.5$~\AA.

The density of the CH$_4$ system was compared to the experimental value~\cite{harris1980density} of 0.4256~g/cm$^3$ (Figure~\ref{fig:densities}~(c), Table~\ref{tab:densities}). The errors obtained for all MLIPs are small and close in magnitude. Furthermore, no significant differences in the accuracy of MLIP(l)'s and MLIP+D's are observed similarly to the modeling of CCl$_4$. 
In contrast, the density of toluene predicted by ETN exhibits a large uncertainty, indicating substantial deviations in the MLIP training results (Figure~\ref{fig:densities}~(d), Table~\ref{tab:densities}). Furthermore, the mean density evaluated with ETN significantly overestimates the experimental value~\cite{harris2000temperature} of 0.862~g/cm$^3$. For the ETN + D2 potential, these issues are resolved, and a density value closer to the experiment is achieved together with a lower spread. MTP also shows higher error and uncertainty compared to MTP+D2, although the difference is not as dramatic as for ETN and ETN+D2. 

To sum up, we observed that the MLIP+D's were more accurate for density calculations than MLIP(l)'s for CCl$_4$ when the cutoff radii were set to 5.5~\AA. However, when the cutoff radii of MLIP(l)'s were extended, their accuracy in modeling CCl$_4$ and CH$_4$ improved to the level of MLIP+D's and even surpassed MTP+D2 for MTP potentials. In contrast, for toluene, ETN+D2 significantly outperformed ETN even when its cutoff radius was extended. This is likely attributed to the unphysical oscillations observed in toluene binding curves (Figure \ref{fig:tol_curves}), which were significantly reduced after incorporating the dispersion correction.

\subsection{Radial distribution functions}

\begin{figure*}[!ht]
    \centering
    \includegraphics[width=1\linewidth]{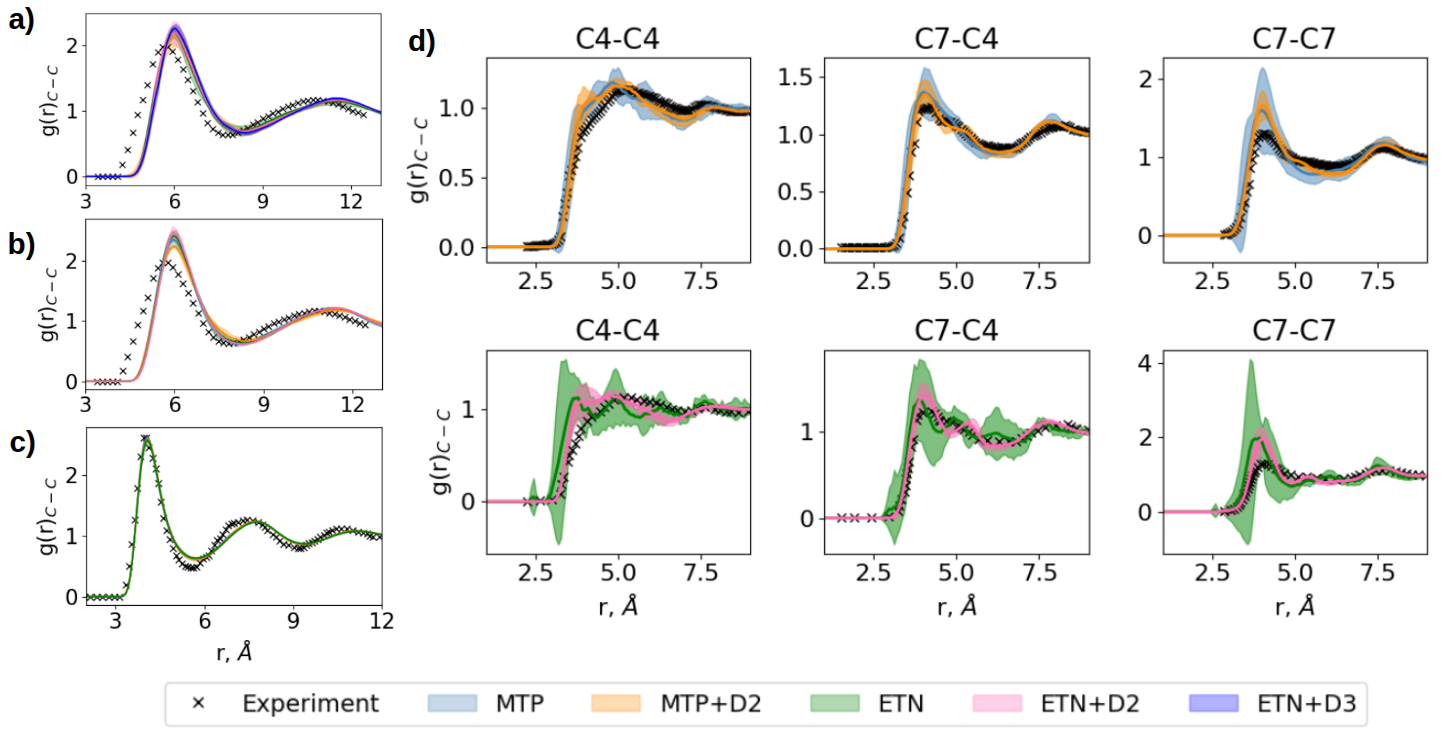}
    \caption{Radial distribution functions obtained from experiment~\cite{lowden1974theory,oobatake1990molecular,falkowska2016neutron} and with MLIPs for: a) CCl$_4$ using MLIPs with cutoff 5.5~\AA; b) CCl$_4$ using MLIP(l)'s with cutoff 7.5~\AA\ and MLIP+D's with cutoff 6~\AA; c) CH$_4$ using MLIP(l)'s with cutoff 7.5~\AA\ and MLIP+D's with cutoff 6~\AA; d) toluene using MLIP(l)'s with cutoff 7.5~\AA\ and MLIP+D's with cutoff 6~\AA.}
    \label{fig:RDFs}
\end{figure*}

Finally, we calculated RDFs, which were obtained by averaging over 100~ps of the NVE MD trajectories after the equilibration of the density in the NpT ensemble and the temperature in the NVT ensemble.
The details of the simulations can be found in Section \ref{comptational_details}. The RDFs calculated with MLIPs are shown in Figure \ref{fig:RDFs} and Table \ref{tab:RDFs} contains the MAPEs for extrema positions and intensities, compared to those obtained experimentally for CCl$_4$~\cite{lowden1974theory}, CH$_4$~\cite{oobatake1990molecular}, and toluene~\cite{falkowska2016neutron}.

\begin{table}
    \centering
    \begin{tabular}{c|c|c|c|c|c|c}
        \hline
        \hline
          & Error& MTP & MTP+D2 & ETN & ETN+D2 & ETN+D3\\ \hline
         \multirow{4}{1cm}{CCl$_4^{(a)}$}&\multirow{2}{*}{MAPE(g)} & \multirow{2}{*}{-} & 0.04& 0.06& 0.06& 0.06\\
         & & & $\pm$ 0.03& $\pm$ 0.04& $\pm$ 0.05& $\pm$ 0.05\\ \cline{2-7}
         &\multirow{2}{*}{MAPE(r)} & \multirow{2}{*}{-} & 0.04& 0.03& 0.03& 0.03\\ 
         & & & $\pm$ 0.02& $\pm$ 0.02& $\pm$ 0.02& $\pm$ 0.02\\ \hline \hline
          \multirow{4}{1cm}{CCl$_4^{(b)}$}&\multirow{2}{*}{MAPE(g)} & 0.10& 0.06& 0.12& 0.13& \multirow{2}{*}{-} \\
         & & $\pm$ 0.08&  $\pm$ 0.05& $\pm$ 0.09& $\pm$ 0.09& \\ \cline{2-7}
         & \multirow{2}{*}{MAPE(r)} & 0.03& 0.03& 0.03& 0.03& \multirow{2}{*}{-} \\ 
         & & $\pm$ 0.01& $\pm$ 0.02& $\pm$ 0.01& $\pm$ 0.01&  \\ \hline \hline
          \multirow{4}{*}{CH4} &\multirow{2}{*}{MAPE(g)} & 0.04& 0.03& 0.04& 0.03& \multirow{2}{*}{-}\\
          & & $\pm$ 0.02& $\pm$ 0.01& $\pm$ 0.01&$\pm$ 0.01& \\ \cline{2-7}
         &\multirow{2}{*}{MAPE(r)}\ & 0.02& 0.02& 0.02& 0.03& \multirow{2}{*}{-}\\ 
         & & $\pm$ 0.02& $\pm$ 0.01& $\pm$ 0.02&$\pm$ 0.02& \\ \hline \hline
    \end{tabular}
    \caption{MAPE and two-$\sigma$ confidence interval of positions and intensities of RDF maxima, compared to experiment for CCl$_4$~\cite{lowden1974theory} and CH$_4$~\cite{oobatake1990molecular}. Where CCl$_4^{(a)}$ refers to results obtained using MLIPs with cutoff 5.5~\AA, while CCl$_4^{(b)}$ refers to results obtained using MLIP(l)'s with cutoff 7.5~\AA\ and MLIP+D's with cutoff 6~\AA.}
    \label{tab:RDFs}
\end{table}

All MLIPs with $R_{\rm cut}=5.5$~\AA\ give nearly equivalent RDFs for CCl$_4$ (Figure \ref{fig:RDFs} (a), Table \ref{tab:RDFs}),  at the same time exhibiting a shift compared to the experimental data due to an underestimation of the density. In contrast to the results obtained for the density, we do not observe a substantial improvement in accuracy after incorporating dispersion corrections in the present case. Note that the RDF was not calculated with MTP as it failed to maintain the liquid phase in the MD simulations. When CCl$_4$ is modeled with MLIP(l)'s with a cutoff of 7.5~\AA, and MLIP+D's with cutoff 6~\AA, the predictions of all MLIPs are close to each other (Figure \ref{fig:RDFs}~(b), Table \ref{tab:RDFs}), similarly to the calculations using MLIPs with smaller cutoffs.

For the CH$_4$ system, similarly to the accurate density predictions, the resulting RDFs are close to the experimental ones and are nearly the same for all (see Figure \ref{fig:RDFs}~(c), Table \ref{tab:RDFs}). Additionally, we note that the pressure conditions of the simulations for density and RDF calculations differ significantly from those during the MD simulations in the AL process, yet the results are highly accurate, indicating excellent transferability of the trained MLIPs.

For modeling toluene, the configurations were collected every 50~fs and post-processed using MDAnalysis~\cite{michaud2011mdanalysis}.
The RDFs were calculated based on the C4 carbon atoms (opposite to the methyl group) and the C7 carbon atoms (methyl carbon).
Three types of RDFs were computed: C7--C4, C4--C4, and C7--C7 (Figure \ref{fig:RDFs}~(d)).
Note that in the C7--C4 RDF the intramolecular contribution was subtracted.

RDFs obtained with all the MLIPs, except for ETN, show qualitative agreement with the experiment. In contrast, the average C4--C4 RDF predicted with ETN has a false maximum. In addition, individual RDFs predicted by individual ETN potentials in the ensemble demonstrate significant spread from each other, leading to broad uncertainty regions. For the ETN+D2 potential, the width of the uncertainty region is substantially reduced and a better agreement with the experiment is achieved. The difference between MTP and MTP+D2 is less pronounced than for ETN and ETN+D2. However, a reduction in the uncertainty region is still observed, along with an improved alignment between the mean and experimental RDFs. As in the case of density prediction, we associate the higher accuracy of the MLIP+D's with reduced oscillations observed on the binding curves (Figure \ref{fig:tol_curves}).

Overall, we observed that the CCl$_4$ and CH$_4$ RDFs were not sensitive to the incorporation of dispersion corrections in the functional form of the MLIPs. However, it had a significant effect on the RDF for toluene, highlighting that this effect depends on system complexity.

Additionally, we want to emphasize that the MLIP+D models still have advantages for the extended-cutoff MLIP(l)'s. In fact, employing a smaller $R_{\rm cut}$ while incorporating explicit dispersion corrections improves the computational efficiency of potential fitting and subsequent simulations. This efficiency stems from the reduced computational cost of DFT calculations for smaller configurations, as well as the faster training and evaluation of energies and forces with potentials that have a smaller cutoff radius.

It should be also noted that we started this study by calculating binding energies, density, and RDF using MLIPs with $R_{\rm cut}=5.5$~\AA ~for the CCl$_4$ system. We found that ETN+D3 outperformed ETN+D2 only for the CCl$_4$ binding curves and, for this reason, we did not use ETN+D3 for other tests provided here.

Finally, the remaining deviations between the predicted densities and RDFs and experimental ones arises not only from MLIP fitting errors but also from intrinsic limitations of the PBE density functional, the dispersion correction schemes, and the neglect of many-body and quantum nuclear effects. While addressing these factors could potentially enhance the accuracy compared to experiment, it was beyond the scope of the present study.

\section{Conclusions}

In this work, we considered the problem of using MLIPs to model systems with strong long-range dispersion interactions. A typical MLIP cutoff radius is 5--6~\AA\ which is much less than the range of dispersion interactions, hence the idea to include explicit dispersion interactions on top of a general-purpose MLIP.

We studied the impact of explicit incorporation of dispersion interactions in MTPs and ETN potentials via the D2 and D3 correction schemes. The impact was measured based on training and validation errors, errors in predicting binding curves, density, and RDFs of a liquid phase of three systems -- carbon tetrachloride (CCl$_4$), methane (CH$_4$), and toluene.

Our first finding is that the accuracy of a typical MLIP (with a cutoff radius of 5.5~\AA) is significantly improved after incorporating explicit dispersion. We observe an improvement in accuracy for predicting energy/forces and binding curves, and this improvement consistently translates into a higher accuracy on the finite-temperature properties. 

Comparing the D2 and D3 corrections, we found that the incorporation of D2 is sufficient, as it provides the same order of accuracy as D3 in the systems we tested. This finding suggests that D2 can be a practical choice for many applications, as D3 is more computationally expensive (or at least harder to implement efficiently).

Another finding is that for the two relatively simple systems, CCl$_4$ and CH$_4$, the improvement in accuracy can instead be achieved by simply increasing the cutoff radius of the local MTP and ETN to 7.5~\AA, except for the full-width at the half-minimum metrics in CH$_4$, in which the MLIP+D's are still more accurate. 
We note that we did not systematically study the effect of increasing the cutoff radius beyond 7.5~\AA\ (due to the fast increase in the cost of collecting the appropriate training set). For the toluene system, which has a sufficiently complex geometry, a simple increase in the cutoff radius was not enough to achieve the accuracy of the MLIP+D models, especially for the ETN model.

\section*{Supplementary material}
See the supplementary material for additional data and supporting discussion of statements made in the main text.  

\begin{acknowledgments}
This work was in part supported by the Russian Science Foundation (grant number 23-13-00332, https://rscf.ru/project/23-13-00332/).

Max Hodapp acknowledges the financial support under the scope of the COMET program within the K2 Center “Integrated Computational Material, Process and Product Engineering (IC-MPPE)” (Project No 886385); this program is supported by the Austrian Federal Ministries for Climate Action, Environment, Energy, Mobility, Innovation and Technology (BMK) and for Labour and Economy (BMAW), represented by the Austrian Research Promotion Agency (FFG), and the federal states of Styria, Upper Austria and Tyrol.
\end{acknowledgments}

\section*{Author declarations}

\subsection*{Conflict of interest}

The authors have no conflicts to disclose.

\subsection*{Author Contributions}

\textbf{Olga Chalykh}: Conceptualization (equal); Methodology (equal); Software (equal); Formal analysis; Writing - original draft; Writing - review \& editing (supporting). \textbf{Dmitry Korogod}: Software (equal); Writing - review \& editing (supporting). \textbf{Ivan S. Novikov}: Methodology (equal); Software (supporting); Writing - original draft (supporting); Writing - review \& editing (equal). \textbf{Max Hodapp}: Methodology (equal); Software (supporting); Writing - original draft (supporting); Writing - review \& editing (supporting). \textbf{Nikita Rybin}: Methodology (supporting); Writing - review \& editing (supporting). \textbf{Alexander V. Shapeev}: Conceptualization (equal); Methodology (equal); Writing - review \& editing (equal).

\section*{Data Availability Statement}

The MLIP-2 code used for the fitting and simulations with MTP is publicly available in the GitLab repository at https://gitlab.com/ashapeev/mlip-2.

The MLIP-4 code used for the fitting and simulations with ETN is publicly available in the GitLab repository at https://gitlab.com/ashapeev/mlip-4.

\bibliography{references}

\newpage
\setcounter{equation}{0}
\setcounter{figure}{0}
\setcounter{table}{0}
\setcounter{page}{1}
\makeatletter
\renewcommand{\theequation}{S\arabic{equation}}
\renewcommand{\thefigure}{S\arabic{figure}}
\renewcommand{\bibnumfmt}[1]{[S#1]}
\renewcommand{\citenumfont}[1]{S#1}

\section{Supplementary material}

\subsection{Accessing the accuracy of the binding curves}

\begin{figure}[!ht]
    \centering
    \includegraphics[width=0.7\linewidth]{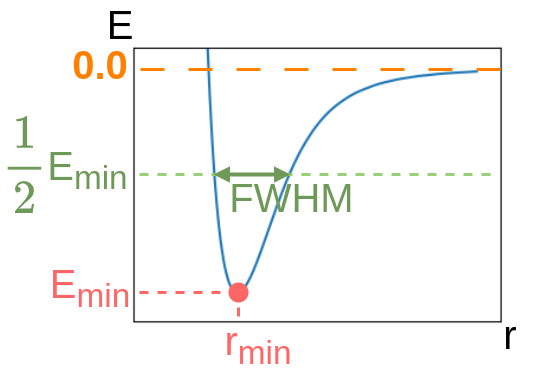}
    \caption{Error metrics proposed for assessing accuracy of MLIPs in predicting dimer binding curves.}
    \label{fig:binding_curves_metrics}
\end{figure}

The metrics for accessing the accuracy of the binding curves predicted by MLIPs: energy and position of the minima along with the full width at the half of the minima (FWHM) are illustrated in Figure \ref{fig:binding_curves_metrics}.

\subsection{Absolute values of minima energies and FWHMs}

\begin{figure*}[!ht]
    \centering
    \includegraphics[width=1\linewidth]{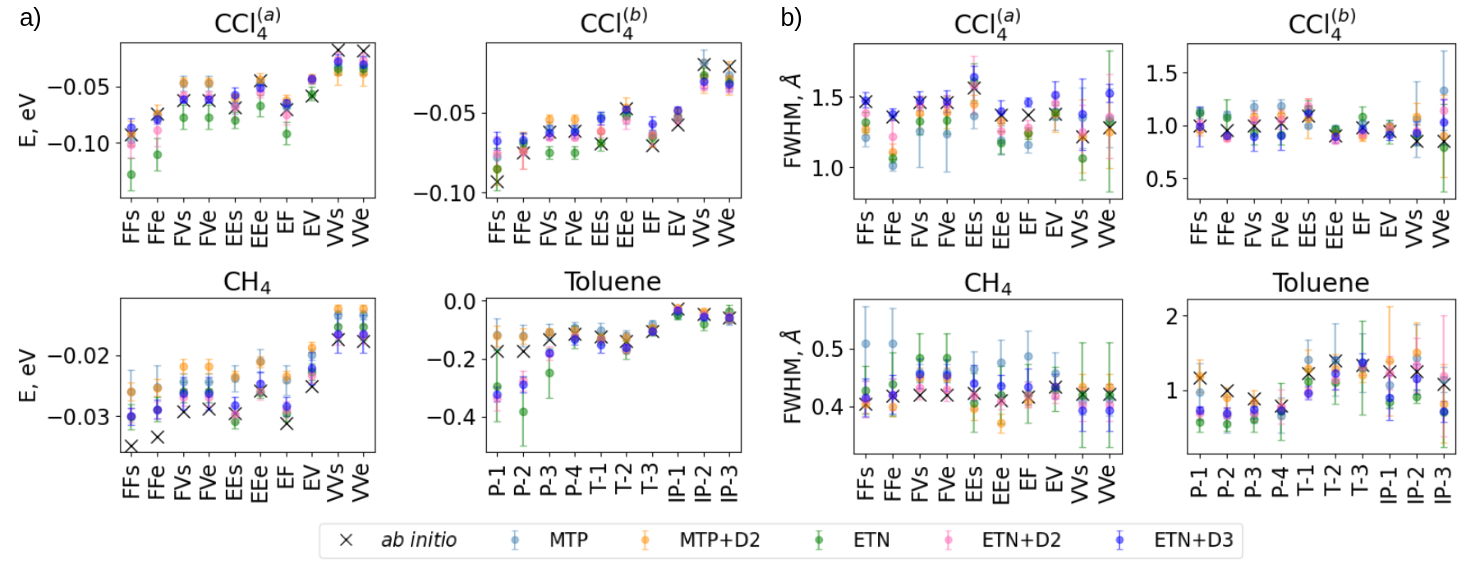}
    \caption{a) Absolute values of E$_{\rm min}$; b) absolute values of FWHM, calculated by \textit{ab initio} method and predicted by MLIPs for: CCl$_4$ using MLIPs with $R_{\rm cut}=5.5$~\AA\ (CCl$_4^{(a)}$), CCl$_4$ using MLIP(l)'s with $R_{\rm cut}=7.5$~\AA\ and MLIP+D's with $R_{\rm cut}=6$~\AA\ (CCl$_4^{(b)}$), CH$_4$ using MLIP(l)'s with $R_{\rm cut}=7.5$~\AA\ and MLIP+D's with $R_{\rm cut}=6$~\AA, and toluene using MLIP(l)'s with $R_{\rm cut}=7.5$~\AA\ and MLIP+D's with $R_{\rm cut}=6$~\AA.}
    \label{fig:SI_absolute_values}
\end{figure*}

Figure \ref{fig:SI_absolute_values} illustrates the absolute values of the minima energies and FWHMs calculated by \textit{ab initio} method and predicted by MLIPs. This figure demonstrates that the incorporation of dispersion correction significantly impacts these values when CCl$_4$ is modeled by MLIPs with $R_{\rm cut}=5.5$~\AA, mitigating large energy errors of ETN and FWHM errors of MTP. When the cutoff radius of MLIP(l)'s is extended to 7.5~\AA\ and for MLIP+D's $R_{\rm cut}$ is set to 6~\AA, there is no significant difference between MLIP(l)'s and MLIP+D's. In contrast, for CH$_4$ the difference is observed even in this case as FWHM is predicted more accurately by MLIP+D's. Additionally, the impact of incorporation of dispersion correction for toluene is not pronounced.

\subsection{Binding curves for the CCl$_4$ and CH$_4$ dimers}

\begin{figure*}[!ht]
    \centering
    \includegraphics[width=1\linewidth]{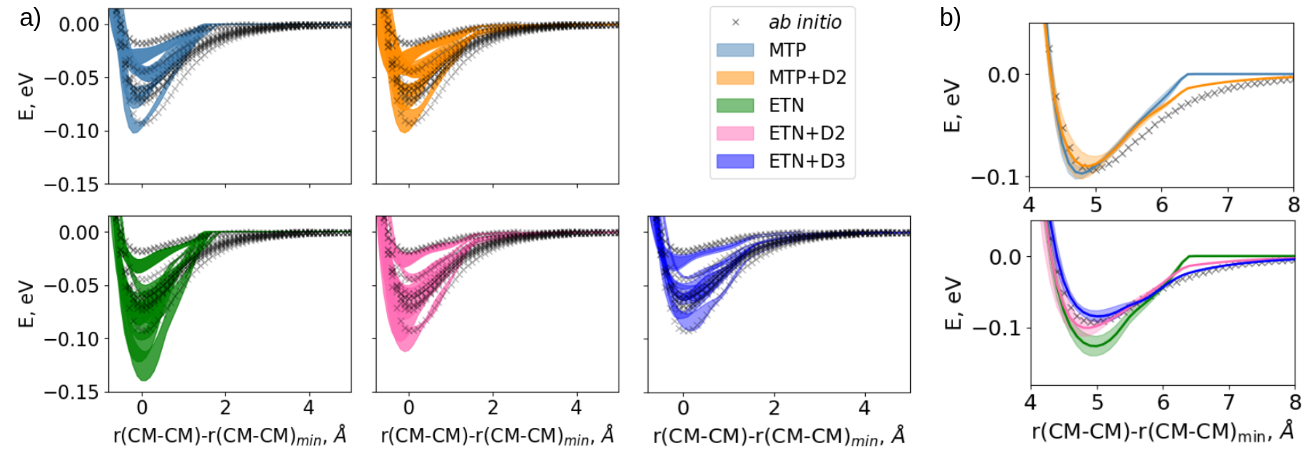}
    \caption{a) All CCl$_4$ binding curves predicted by MLIPs with $R_{\rm cut}=5.5$~\AA; b) FFs dimer binding curve.}
    \label{fig:SI_ccl4_55_curves}
\end{figure*}

Binding curves of CCl$_4$ dimers predicted by MLIPs with $R_{\rm cut}=5.5$~\AA\ are shown in Figure \ref{fig:SI_ccl4_55_curves}. This figure demonstrates that MLIP(l)'s possess inaccurate asymptotics, which is mitigated by incorporation of dispersion correction. Moreover, we found that when CCl$_4$ binding curves are modeled by MLIPs with $R_{\rm cut}=5.5$~\AA, the MLIP(l)'s do not possess the correct $1/r^6$ asymptotic behavior, as outside the cutoff MLIP(l)'s treat atoms as noninteracting. The asymptotic behavior slightly improves when dispersion interactions are included via D2, however there is an observable difference compared to the ab initio binding curve, likely due to the difference between D2 $C_6$ coefficients used in the dispersion correction of MLIPs and D3 $C_6$ coefficients used in ab initio calculations. The best asymptotic behavior is achieved by ETN+D3.

\begin{figure*}[!ht]
    \centering
    \includegraphics[width=1\linewidth]{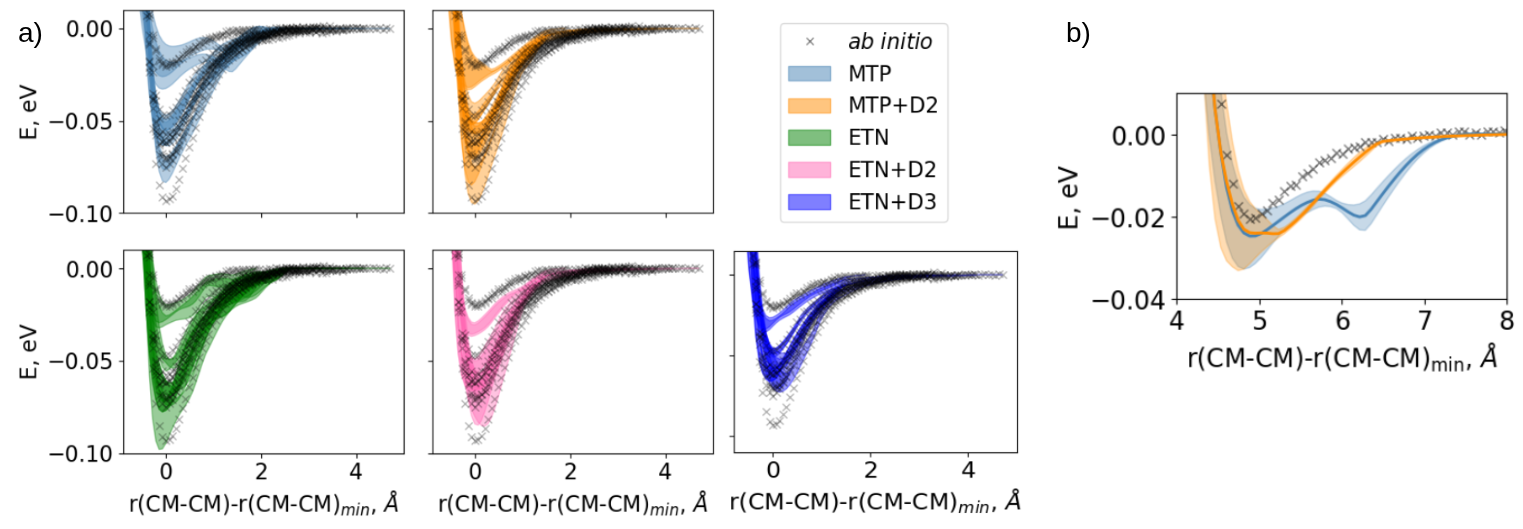}
    \caption{a) All CCl$_4$ binding curves predicted by MLIP(l)'s with $R_{\rm cut}=7.5$~\AA\ and MLIP+D's with $R_{\rm cut}=6$~\AA; b) VVe dimer binding curve.}
    \label{fig:SI_ccl4_75_curves}
\end{figure*}

When the cutoff radius is extended for MLIP(l)'s to 7.5~\AA, and for MLIP+D's to 6~\AA, MLIP(l)'s predict binding curves with more accurate asymtotics, and overal reaching the level of accuracy of MLIP+D's (Figure \ref{fig:SI_ccl4_75_curves}~ (a)). However, VVe dimer binding curve predicted by MTP exhibited unphysical oscillations, illustrated in Figure \ref{fig:SI_ccl4_75_curves}~(b). While the magnitude of the oscillation remains small, it is comparable to the VVe dimer binding curve minima depth.  This effect is also mitigated by incorporation of dispersion correction in MTP+D2.

In contrast, binding curves of CH$_4$ dimers modeled by MLIPs with extended cutoffs are free from the unphysical oscillations and are overall more accurate than CCl$_4$ ones.

\begin{figure*}[!ht]
    \centering
    \includegraphics[width=0.7\linewidth]{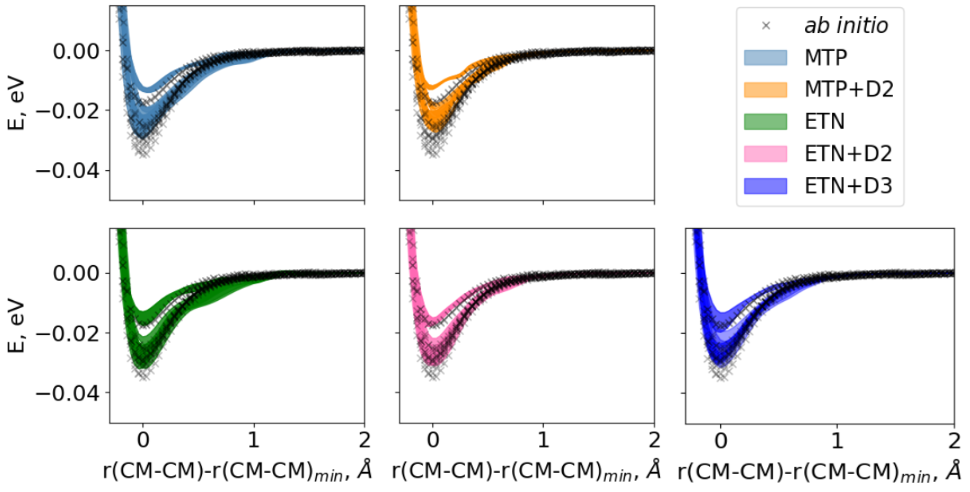}
    \caption{All CH$_4$ binding curves predicted by MLIP(l)'s with $R_{\rm cut}=7.5$~\AA\ and MLIP+D's with $R_{\rm cut}=6$~\AA.}
    \label{fig:SI_ch4_curves}
\end{figure*}

\end{document}